\documentclass[aps,twocolumn,rmp,showpacs,amsfonts,floatfix]{revtex4}
\usepackage{graphicx}
\usepackage{color}
\usepackage{ulem}

\begin{document}

\title{{\normalfont\bfseries\slshape Colloquium}: Spontaneous Magnon Decays}
\author{M. E. Zhitomirsky}
\email{mike.zhitomirsky@cea.fr}
\affiliation{
Service de Physique Statistique, Magn\'etisme et Supraconductivit\'e,
UMR-E9001 \\ CEA-INAC/UJF,  17 rue des Martyrs,
38054 Grenoble cedex 9, France}

\author{A. L. Chernyshev}
\email{sasha@uci.edu}
\affiliation{Department of Physics, University of California, Irvine,
California 92697, USA}

\date{23 January 2013}

\begin{abstract}
A theoretical overview of the phenomenon of spontaneous magnon
decays in quantum antiferromagnets is presented. The intrinsic zero-temperature damping
of magnons  in quantum spin systems is a fascinating many-body effect, which has
recently attracted significant attention in view of its possible observation
in neutron-scattering experiments. An introduction to the
theory of magnon interactions and a discussion of necessary symmetry and kinematic
conditions for spontaneous decays are provided. Various parallels with the decays of
anharmonic phonons and excitations in superfluid $^4$He are extensively used.
Three principal cases of spontaneous magnon decays are considered:
field-induced decays in Heisenberg antiferromagnets, zero-field decays
in spiral antiferromagnets, and triplon decays in quantum-disordered
magnets. Analytical results are compared with available numerical data and
prospective materials for experimental observation of the decay-related
effects are briefly discussed.
\end{abstract}

\maketitle

\tableofcontents

\section{Introduction}
\label{sec:intro}

According to conventional wisdom a quasiparticle is presumed well-defined
until proven not to be. The textbook picture of magnons as long-lived
excitations weakly interacting with each other works very well for many
magnetic materials. Nevertheless,  in a number of recently studied
spin systems magnons neither interact weakly nor remain well-defined
even at zero temperature. This is due to the so-called spontaneous quasiparticle 
decay, a spectacular quantum-mechanical many-body effect, which is the topic of
the current Colloquium.

Magnetic excitations have been the subject of intensive theoretical and
experimental studies since the first half of the last century. Historically,
the concept of spin waves was introduced by \textcite{Bloch30} and,
independently, by \textcite{Slater30} for the wavelike propagation of spin
flips in ordered ferromagnets; see also \textcite{Hoddeson87}.
Already in his earliest work \textcite{Bloch30} described spin waves as weakly
interacting excitations obeying Bose statistics. Subsequently,
\textcite{Holstein40} introduced second quantization of spin waves in terms of
bosonic operators, the approach widely used to this day. The first use of
the word ``magnon,'' which concisely unifies the notions of an elementary
quasiparticle and a magnetic quantum, can be traced back to
\textcite{Pomeranchuk41}, who ascribed this term to Lev Landau, see
\textcite{Walker70}. Although the terms ``spin waves'' and ``magnons'' are
sometimes used to distinguish between long- and short-wavelength excitations,
respectively, in the following we use both names interchangeably.

In addition, the fundamental contribution to the theory of spin waves in
magnetic insulators was made by \textcite{Anderson52},
\textcite{Kubo52}, and  \textcite{Dyson56a,Dyson56b}. These and subsequent studies
developed a comprehensive picture of spin waves in conventional quantum ferromagnets
and antiferromagnets, which can be found in various reviews and books, see
\textcite{Kranendonk58}, \textcite{Akhiezer}, \textcite{Mattis}, \textcite{White},
\textcite{Borovik88}, \textcite{Manousakis91}, and \textcite{Majlis}.
Invented in the 1950s, inelastic neutron-scattering spectroscopy
\cite{Brockhouse95} has quickly become a major experimental
method of investigating spin waves and other magnetic excitations in solids.
With a variety of specialized techniques currently used at numerous neutron
facilities around the world, it remains the most powerful and versatile research
tool for studies of magnetic spectra. Other microscopic techniques to investigate
excitations in magnetic materials are electron spin resonance \cite{Gurevich}
and light scattering methods, such as resonant inelastic X-ray scattering
\cite{Xray}.

Usually, the magnon-magnon interactions play a minor role at low temperatures.
For example, as $T\to 0$ deviation of the spontaneous magnetization of a
ferromagnet from its saturated value follows Bloch's law
$\Delta M \propto T^{3/2}$, the result obtained by considering magnons
as noninteracting bosons. Interactions yield a correction to Bloch's law
that scales as $T^4$, which is still smaller than the ``kinematic'' correction
proportional to $T^{5/2}$ due to deviation of magnon dispersion from its parabolic form
\cite{Dyson56b}. In antiferromagnets, effects of magnon-magnon interaction survive
at zero temperature due to zero-point motion in the ground state. An estimate of
their role is given by the magnitude of the quantum correction to the magnon dispersion
law obtained in the harmonic, or semiclassical, approximation. Even in low
dimension (2D) and even in the ``extreme quantum'' case of $S=1/2$, such a
correction  for the square-lattice Heisenberg antiferromagnet amounts to
a mere $\sim 16$\%\ rescaling of the harmonic dispersion \cite{Manousakis91}.

Recently, there has been a growing body of theoretical and experimental studies
showing that the conventional picture of magnons as weakly interacting quasiparticles
is not always correct. The first class of systems exhibiting significant deviations
from the conventional semiclassical dynamics is weakly coupled antiferromagnetic
chains, for which the spin-wave approximation is not a good starting point.
Their high-energy spectra are well-described by propagating one-dimensional spinons,
the natural basis for excitations in spin-$\case{1}{2}$ chains \cite{Schulz96,Coldea01,Kohno07}.
Since the development of a quantitative theory for the 1D to 3D crossover in weakly
coupled chains is still in progress, we shall not discuss this physics here.

The second class is, in a sense, simpler and more puzzling at the same time.
It is comprised of quantum magnets with well-ordered ground states whose dynamical
properties are yet quite different from the predictions of harmonic spin-wave theory
\cite{Zheng06a}. In the following, we focus on this second class of unconventional
magnetic systems. It includes quantum magnets with noncollinear spin ordering.
The key role in their unusual dynamical properties is played by cubic anharmonicities,
or three-magnon processes, which originate from the noncollinearity of spins.
Three-magnon processes are absent in conventional collinear antiferromagnets,
making their spin-wave excitations intrinsically weakly interacting.
The most striking effect of three-magnon interactions is the finite lifetime of
spin waves: not only at nonzero temperatures when magnon linewidth is determined
by scattering off thermal excitations, but also at $T=0$ when damping
is produced by {\it spontaneous} magnon decay.

In order to put the phenomenon of spontaneous magnon decays in
a broader context, we note that two other subfields of
condensed matter physics also deal with decays of bosonic excitations.
These are anharmonic crystals \cite{Kosevich}
and superfluid liquids and gases \cite{StatPhysII,Pethick}.
Cubic anharmonicities are naturally present in these systems, but their
role differs both quantitatively and qualitatively from that in quantum spin systems.
In crystals, the anharmonic corrections are generally small due to small
mean fluctuations of ions around equilibrium positions, the condition
satisfied for the majority of solids below the Debye temperature.
In superfluids, the strength of cubic interactions is determined
by the condensate fraction and may again become small, as is the case of superfluid $^4$He.
Quantum magnets are unique in that (i) cubic anharmonicities may be tuned on and off
by an external magnetic field or by changing the lattice geometry,
and  (ii)  the existing variety of magnetic compounds
provides diverse forms of the magnon dispersion law $\varepsilon_{\bf k}$,
which plays key role in spontaneous quasiparticle decay.
Still, various parallels and analogies between the three branches of
condensed matter physics prove to be  helpful
and will be extensively used in the subsequent discussion.

The rest of the article is organized as follows.
Section~\ref{sec:anharmonic} contains a general discussion of the origin of cubic as well as quartic
anharmonicities in quantum magnets. The energy and
momentum conservation conditions for spontaneous  decays and  the decay-induced singularities
in two-dimensional spin systems are analyzed in Sec.~\ref{sec:kinematics}.
In Sec.~\ref{sec:field_decays} we provide  a description of spontaneous magnon decays in
the square-lattice Heisenberg antiferromagnet in strong external field,
while Sec.~\ref{sec:tafm} deals with another prototypical noncollinear magnet:
the triangular-lattice Heisenberg antiferromagnet in zero field.
A few general scenarios of the decay of long-wavelength excitations are
considered in Sec.~\ref{sec:damping_lowE}.
Section \ref{sec:spin_liquids} is devoted to the discussion of the role
of cubic anharmonicities in gapped quantum  spin liquids.
Finally, a summary and outlook are presented in Sec.~\ref{sec:summary}.

\section{Magnon-magnon interactions}
\label{sec:anharmonic}

In the classical picture, the interaction between magnons is related to the amplitude
of spin precession and can become arbitrarily small as the amplitude decreases.
Zero-point motion of spins puts a lower bound on the amplitude and leads to
non-negligible interaction effects in quantum magnets. Usually, the consideration
of magnon-magnon interactions begins with a discussion of bosonic representations
for spin operators. Here we intentionally skip this step, postponing technical
details to the subsequent sections. Instead, we discuss the general structure of
effective bosonic Hamiltonians defined solely by the symmetry
of the magnetic ground state and corresponding excitations.

\subsection{Four-magnon interactions}
\label{sec:4mag_sec}

To draw an analogy between different bosonic systems
we begin with the Hamiltonian  of a normal Bose gas
\begin{eqnarray}
\hat{\cal H}_0 & = & \sum_{\bf k} \varepsilon_{\bf k}
a^\dagger_{\bf k} a^{_{}}_{\bf k}
\label{Hfm} \\
& + &  \frac{1}{4} \sum_{{\bf k}_i}  
V^{(0)}_{4}({\bf k}_1,{\bf k}_2;{\bf k}_3,{\bf k}_4)\,
a^\dagger_{{\bf k}_1} a^\dagger_{{\bf k}_2} a^{_{}}_{{\bf k}_3} a^{_{}}_{{\bf k}_4}+\dots ,
\nonumber
\end{eqnarray}
where $\varepsilon_{\bf k}$ is the kinetic energy,
$V^{(0)}_{4}({\bf k}_1,{\bf k}_2;{\bf k}_3,{\bf k}_4)$ is the two-particle scattering
amplitude, the ellipsis stands for $n$-particle interactions, and momentum conservation
is assumed from now on for various $\bf k$-summations. The interaction term in Eq.~(\ref{Hfm})
conserves the number of particles, a natural constraint for liquid helium or cold atomic
gases. At the same time, an effective bosonic Hamiltonian with exactly the same
structure as in Eq.~(\ref{Hfm}) describes the quantum Heisenberg ferromagnet
\cite{Holstein40,Oguchi60}. Although the number of quasiparticles
may not be conserved, in a ferromagnet the conservation is enforced by the invariance
of the ground state under an arbitrary rotation about the magnetization direction.
Because of this symmetry the ground state and all excitations are characterized by definite
values of the $z$-projection of the total spin $S^z$, from $S^z_{\rm tot} = NS$ and down.
Consequently, the matrix elements of the Hamiltonian vanish for states with different
spin projections, resulting in the particlelike form of the interaction term in Eq.~(\ref{Hfm}).
In other words, in the Heisenberg ferromagnet
every magnon has an intrinsic quantum number $\Delta S^z=-1$,
which is conserved in magnon scattering processes.

The quasiparticle-number conservation does not hold for interaction processes
in quantum antiferromagnets. As pointed out by \textcite{Anderson},
the symmetry-broken antiferromagnetic ground state corresponds to
a superposition of states with different values of the total spin.
Hence, elementary excitations cannot be assigned with
a definite value of spin: the spin of the spin wave ceases to exist.  Because of that,
the effective Hamiltonian of a quantum antiferromagnet contains
additional interaction terms,
\begin{eqnarray}
&& \hat{\cal H}_4  =
\frac{1}{3!}\! \sum_{{\bf k}_i} V^{(1)}_{4}\!({\bf k}_1,{\bf k}_2,{\bf k}_3;{\bf k}_4)\,\bigl(
a^\dagger_{{\bf k}_1} a^\dagger_{{\bf k}_2} a^\dagger_{{\bf k}_3}
a^{_{}}_{{\bf k}_4}\! + {\rm h.c.}\bigr)
\nonumber \\
&& +  \frac{1}{4!} \sum_{{\bf k}_i}
V^{(2)}_{4}\!({\bf k}_1,{\bf k}_2,{\bf k}_3,{\bf k}_4)\,\bigl(
a^\dagger_{{\bf k}_1} a^\dagger_{{\bf k}_2} a^\dagger_{{\bf k}_3} a^\dagger_{{\bf k}_4}
\!+ \textrm{h.c.}\bigr),
 \label{Haf}
\end{eqnarray}
that do not conserve the number of excitations; see, for example, \textcite{Harris71}.
The first term in Eq.~(\ref{Haf})  describes decay and recombination processes of
one magnon into three and vice versa,
whereas the second, so-called source term,
corresponds to creation and annihilation of
four particles out of (into) an antiferromagnetic vacuum.

\subsection{Three-magnon interactions}
\label{sec:3mag_sec}

Another type of anharmonicity that may appear in an effective bosonic Hamiltonian
is the three-particle interaction term.
In a generic quantum system with nonconserved number of particles
anharmonicities of all orders are present,
beginning with the lowest-order cubic terms, which couple
one- and two-particle states. For instance, cubic terms represent
the dominant anharmonicity for lattice vibrations
\cite{Kosevich}.
However, additional symmetry restrictions may forbid such cubic interactions
in antiferromagnets.

In the case of a collinear antiferromagnet,
the remaining SO(2) rotational symmetry about the N\'eel vector direction
($z$ axis) prohibits cubic terms. Although the single-magnon state
is not an eigenstate of the $S^z_{\rm tot}$ operator, such a state still preserves
an odd parity under the $\pi$ rotation about the $z$ axis.
In contrast, the two-magnon states are
invariant under this symmetry operation. Consequently, the
coupling between one- and two-particle sectors is strictly forbidden
in collinear antiferromagnets and cubic terms do not occur in their effective bosonic description.
In hindsight, this lack of low-degree anharmonicities explains
why the spin-wave theory works quantitatively well even for
spin-$\frac{1}{2}$ systems: magnets with collinear spin structures
have excitations that are intrinsically weakly-interacting.

On the other hand, three-magnon interaction terms must be present if
the spin-rotational symmetry is completely broken in the ground state,
{\it i.e.}, when the magnetic structure becomes noncollinear. This is realized,
for example, due to spin canting in an applied magnetic field or because
of competing interactions in frustrated antiferromagnets.
The general form of the three-boson interaction is given by
\begin{eqnarray}
&& \hat{\cal H}_3  =
\frac{1}{2!} \sum_{{\bf k}_i}
V^{(1)}_{3}({\bf k}_1,{\bf k}_2;{\bf k}_3)\,\bigl(
a^\dagger_{{\bf k}_1} a^\dagger_{{\bf k}_2} a^{_{}}_{{\bf k}_3}
 + {\rm h.c.}\bigr)
 \nonumber \\
& & 
\phantom{\hat{V}_3=} + \frac{1}{3!} \sum_{{\bf k}_i}
V^{(2)}_{3}({\bf k}_1,{\bf k}_2,{\bf k}_3)\,\bigl(
a^\dagger_{{\bf k}_1} a^\dagger_{{\bf k}_2} a^\dagger_{{\bf k}_3}
+ \textrm{h.c.}\bigr), \quad
\label{V3}
\end{eqnarray}
where the first term describes the two-particle decay and recombination
processes while the second is another source term.

There exists a deep analogy between the noncollinear quantum magnets
and superfluid boson systems regarding the presence of cubic
interactions in their effective Hamiltonians.
The Bose gas in the normal state is invariant under the group of gauge
transformations U(1) and is described by the Hamiltonian (\ref{Hfm}).
The macroscopic occupation of the lowest-energy state
$\langle a_{{\bf k}=0}\rangle \neq 0$ below the superfluid transition
leads to breaking of the U(1) symmetry and, at the same time, to
the appearance of  the cubic terms by virtue of the Bogolyubov substitution
\begin{equation}
a^\dagger_{{\bf k}_1} a^\dagger_{{\bf k}_2} a_{{\bf k}_3} a_{{\bf k}_4}
\rightarrow a^\dagger_{{\bf k}_1} a^\dagger_{{\bf k}_2} a^{_{}}_{{\bf k}_3}
\langle a^{_{}}_0\rangle  + \ldots\, .
\label{condensate}
\end{equation}
In other words, the vacuum state with the Bose condensate absorbs or emits
an extra particle, enabling nonconserving processes between  excitations.

An analogous consideration of the noncollinear magnetically ordered state
is possible, leading to the same qualitative answer. In fact, it was
realized a long time ago by \textcite{Matsubara56} and  by \textcite{Batyev84}
that there is a one-to-one correspondence between breaking the SO(2)
rotational invariance about the field direction for
Heisenberg and planar magnets and breaking the U(1) gauge symmetry
at  the superfluid transition.  Noncollinear magnetic structures
stabilized by competing exchange interactions in zero field belong to a different
class  as they spontaneously break the full SO(3) rotational symmetry
 without the help of an external field. Still,
the absence of any remaining symmetry
constraints for their elementary excitations, apart from energy and momentum conservation,
permits all possible anharmonic terms including the cubic ones.

It is necessary to mention that cubic anharmonicities due to the long-range dipolar
interactions have been discussed already in the early works on the quantum theory of
spin waves in ferromagnets \cite{Holstein40,Akhiezer} where
they play an important role in the low-frequency magnetization dynamics \cite{Gurevich,Chernyshev12}.
However, the effect of cubic terms is completely negligible at higher energies
because of the smallness of the dipole-dipole interaction compared to the exchange energy.
In contrast, cubic anharmonicities discussed in this work are of exchange origin
and have a profound effect on the magnon excitation spectra.

Anharmonic interactions are not restricted to cubic and
quartic processes. The usual spin-wave expansion based on
the Holstein-Primakoff transformation for spin operators produces an
infinite series of interaction terms. Typically, these higher-orders terms
are small due to the spin-wave expansion parameter
$\langle a^\dagger a\rangle/2S\ll 1$ associated with each order of expansion.
We disregard them in the following analysis as they do not alter our results, either qualitatively or
quantitatively.
\begin{figure}[t]
\centerline{
\includegraphics[width=0.95\columnwidth]{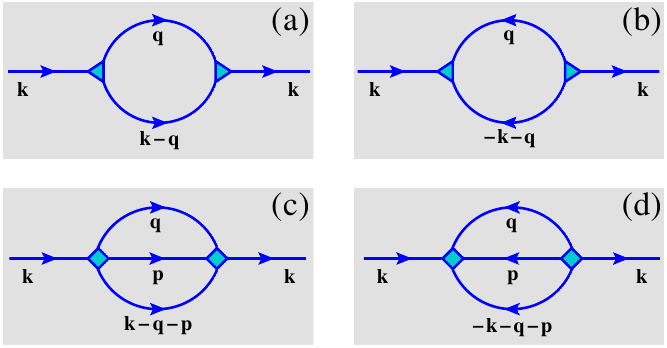}
}
\caption{(color online).
Zero-temperature self-energy diagrams due to the  (a) and (b) three-magnon,
and (c) and (d) four-magnon interactions.
}
\label{fig:diagrams}
\end{figure}

\subsection{Magnon self-energies}
\label{sec:diagrams_sec}

Magnon interactions play a dual role. On the one hand, they renormalize magnon
energy and, on the other hand, they can lead to magnon decay and result in a finite
lifetime. Both effects are treated on equal footing within the standard Green's
function approach \cite{StatPhysII,Mahan}. The lowest-order self-energy diagrams
produced by three- and four-particle vertices at $T=0$ are shown in
Fig.~\ref{fig:diagrams}. Analytic expressions for the diagrams in Figs.~\ref{fig:diagrams}(a) and \ref{fig:diagrams}(b) are
\begin{eqnarray}
\Sigma_a({\bf k},\omega)  & = & \frac{1}{2}\sum_{\bf q}
\frac{\big|V^{(1)}_{3}({\bf q};{\bf k})\big|^2}
{\omega - \varepsilon_{\bf q} \!- \varepsilon_{\bf k-q}\!+i0}\,,
\label{SelfE3}
\\
\Sigma_b({\bf k},\omega) &  = & -\frac{1}{2}\sum_{\bf q}
\frac{\big|V^{(2)}_{3}({\bf q},{\bf k})\big|^2}
{\omega + \varepsilon_{\bf q} + \varepsilon_{\bf k+q}}\,,
\nonumber
\end{eqnarray}
whereas the three-magnon decay process of Fig.~\ref{fig:diagrams}(c) gives
\begin{equation}
\Sigma_c({\bf k},\omega)  =  \frac{1}{6}\sum_{\bf q,p}
\frac{\big|V_4^{(1)}({\bf q}, {\bf p};{\bf k})\big|^2}
{\omega - \varepsilon_{\bf q} - \varepsilon_{\bf p} - \varepsilon_{\bf k-q-p}+i0} \ .
\label{SelfE4}
\end{equation}
We included only independent momenta in the arguments of vertices in Eqs.~(\ref{SelfE3})
and (\ref{SelfE4}).

In the lowest order of perturbation theory, the real part of
the on-shell self-energy $\Sigma({\bf k},\omega\!=\!\varepsilon_{\bf k})$ gives a
correction to the dispersion $\delta\varepsilon_{\bf k}$,
and the imaginary part yields the decay rate
$\Gamma_{\bf k}= -\textrm{Im}\,\Sigma({\bf k},\varepsilon_{\bf k})$.
The source diagrams in Figs.~\ref{fig:diagrams}(b) and \ref{fig:diagrams}(d) have
no imaginary parts, while the magnon damping resulting
from the decay diagram in Fig.~\ref{fig:diagrams}(a) is given by
\begin{equation}
\Gamma^{(3)}_{{\bf k}}   =
\frac{\pi}{2}\sum_{\bf q} \big|V_{3}^{(1)}({\bf q};{\bf k})\big|^2
\delta(\varepsilon_{\bf k} - \varepsilon_{\bf q} - \varepsilon_{\bf k-q}) \ .
\label{Gamma2_k}
\end{equation}
The three-particle decay process in Fig.~\ref{fig:diagrams}(c) yields a similar expression
\begin{equation}
\Gamma^{(4)}_{{\bf k}}   =
\frac{\pi}{6}\sum_{\bf q} \big|V_{4}^{(1)}({\bf q},{\bf p};{\bf k})\big|^2
\delta(\varepsilon_{\bf k} - \varepsilon_{\bf q} - \varepsilon_{\bf p}
- \varepsilon_{\bf k-q-p}) \ .
\label{Gamma3_k}
\end{equation}

Existence of the decay amplitudes $V^{(1)}_{3}$ and $V^{(1)}_{4}$ is a necessary,
but by no means sufficient condition for spontaneous magnon decays.
The decay rate $\Gamma_{\bf k}$ is nonzero only  if
the energy conservation is satisfied for at least one decay channel.
A general analysis of the kinematic conditions following from the energy conservation
for two-particle decays is given in Sec.~\ref{sec:kinematics}.

\subsection{Three-boson interactions in spin liquids}
\label{sec:int_SL}

Apart from magnetically ordered antiferromagnets, there is a wide class of
quantum-disordered magnets with spin-liquid-like ground states. At zero
temperature such magnets remain completely isotropic under spin rotations
and their ground-state wave function is a singlet state of the total spin
$S_{\rm tot}=0$; see Sec.~\ref{sec:spin_liquids} for more detailed discussion
and references. In spin-dimer systems and in some gapped chain materials,
the low-energy excitations are $S=1$ magnons separated by a finite gap
from the singlet ground-state. This means that at any given momentum $\bf k$
there is a triplet of excited states with the same energy $\varepsilon_{\bf k}$.
We use bosonic operators $t^\dagger_{\bf k\alpha}$ and $t_{\bf k\alpha}$ with
$\alpha = x,y$, and $z$ to describe creation or annihilation of such quasiparticles,
which are also called triplons. Having an intrinsic quantum number, spin
polarization $\alpha$, does not change appreciably the structure of the quartic
terms in Eqs.~(\ref{Hfm}) and (\ref{Haf}): the SO(3)-invariant anharmonic
interactions are constructed by making all possible convolutions for two pairs
of polarization indices. An interesting observation is that in this case the
spin-rotational symmetry {\it does not} forbid the cubic vertices. The total
spin is conserved since two spin-1 magnons can form a state with the same
total spin $S_{\rm tot}=1$. The invariant form of the cubic interaction for
triplons is then given by
\begin{equation}
\hat{\cal H}_3\! =
\frac{1}{2!} \sum_{{\bf k}_i}
V_{3}({\bf k}_1,{\bf k}_2;{\bf k}_3)\,\epsilon^{\alpha\beta\gamma}\,\bigl(
t^\dagger_{{\bf k}_1\alpha} t^\dagger_{{\bf k}_2\beta} t^{_{}}_{{\bf k}_3\gamma}\!
 + {\rm h.c.}\bigr),
\label{V3triplon}
\end{equation}
where the conservation of total spin is imposed by the antisymmetric tensor
$\epsilon^{\alpha\beta\gamma}$. In addition, there are certain lattice
symmetries affecting the structure of the decay vertex in Eq.~(\ref{V3triplon}),
which may still forbid the two-particle decays of triplons. Specific examples
of that are discussed in Sec.~\ref{sec:spin_liquids}.

\section{Kinematics of two-magnon decays}
\label{sec:kinematics}

The aim of this section is to consider kinematic constraints in the
two-particle decay process that follow from energy conservation:
\begin{equation}
\varepsilon_{\bf k} = \varepsilon_{\bf q} + \varepsilon_{\bf k-q} \ .
\label{energy_conserv}
\end{equation}
For a magnon with the momentum ${\bf k}$, the relation (\ref{energy_conserv})
is an equation for an unknown ${\bf q}$ with $\bf k$ being an external parameter.
Solutions of Eq.~(\ref{energy_conserv}) form a decay {\it surface}, the locus of
momenta of quasiparticles created in the decay. As a function of the initial
momentum ${\bf k}$ the decay surface expands or shrinks and may disappear
completely. In the latter case, the magnon with momentum ${\bf k}$ becomes
stable. The region in ${\bf k}$ space with stable excitations is separated
from the decay {\it region} where decays are allowed by the  decay
{\it threshold boundary}.

For any given ${\bf k}$ the two-particle excitations form a continuum of
states in a certain energy interval
\begin{equation}
E_2^{\min}({\bf k}) \leq E_2({\bf k},{\bf q}) \equiv \varepsilon_{\bf q} +
\varepsilon_{\bf k-q} \leq  E_2^{\max}({\bf k}) \ .
\label{continuum}
\end{equation}
With the help of this statement it is straightforward to see that nontrivial
solutions of Eq.~(\ref{energy_conserv}) exist only if the single-particle branch and
the two-particle continuum overlap. From this perspective it also becomes evident
that the decay threshold boundary must be the surface of intersections of the
single-particle branch $\varepsilon_{\bf k}$ with the {\it bottom} of the continuum
$E^{\min}_2(\bf k)$.

Suppose for a moment that spontaneous two-particle decays are completely forbidden,
{\it i.e.}, the one-magnon branch lies below the two-magnon continuum
in the whole Brillouin zone
\begin{equation}
\varepsilon_{\bf k} \leq \varepsilon_{\bf q} + \varepsilon_{\bf k-q} \qquad
\textrm{for} \quad \forall\  {\bf k,q} \ ,
\label{below_continuum}
\end{equation}
where the equality may be realized only for a trivial solution ${\bf q =0}$,
provided that $\varepsilon_{0}=0$. Applying the same relation (\ref{below_continuum})
to $\varepsilon_{\bf k-q}$ on the right-hand side gives
\begin{equation}
\varepsilon_{\bf k} \leq \varepsilon_{\bf q} + \varepsilon_{\bf p} + \varepsilon_{\bf k-q-p}
\qquad  \textrm{for} \quad  \forall\  {\bf q,p} \ ,
\end{equation}
which means that the one-magnon branch also lies below the three-magnon continuum
as well as any $n$-magnon continua.
Thus, if two-particle decays of Eq.~(\ref{energy_conserv}) are prohibited for any value of
$\bf k$, the energy and momentum conservation also forbid decays in all
$n$-particle channels with $n\geq 3$ \cite{Harris71}. This is why it is
important to analyze kinematic conditions for two-particle decays as
the first step even if the cubic terms are not present.

Aside from finding whether spontaneous decays exist or not, there is another
reason for considering kinematics of decays. Investigating
two-roton decays in the superfluid $^4$He, \textcite{Pitaevskii59} found that
the enhanced density of states (DoS) near the bottom of the two-particle continuum
may produce strong singularities in the single-particle spectrum at the decay
threshold boundary.
Additional singularities may also occur inside
the decay region due to topological transitions of the decay surface
\cite{Chernyshev06}. A general analysis of these two effects is given in
Sec.~\ref{sec:kinematics_singular}.

\subsection{Decay threshold boundary}
\label{sec:kinematics_threshold}

\begin{figure}[t]
\includegraphics[width=0.75\columnwidth]{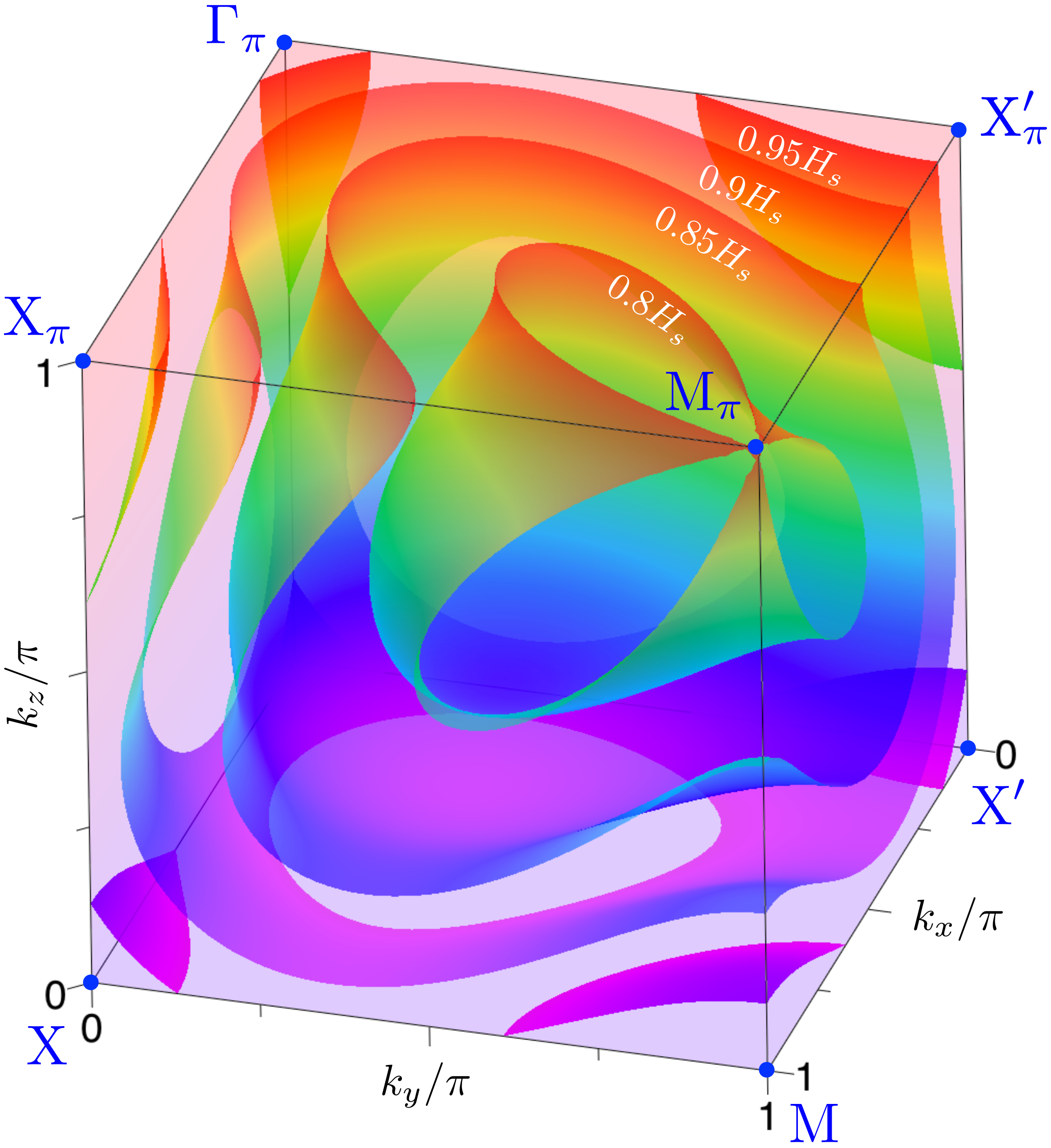}
\caption{(color online).  Magnon decay threshold boundaries for the cubic
antiferromagnet in an external field shown in one-eighth of the Brillouin zone for
several values of the field. Decays are possible within the part of
the Brillouin zone enclosing the M$_\pi$ point.
}
\label{regions}
\end{figure}

The bottom of the two-particle continuum can be found
by imposing  the extremum condition $\nabla_{\bf q} E_2({\bf k},{\bf q}) = 0$,
which is satisfied if the  velocities of two magnons are equal,
\begin{equation}
{\bf v}_{\bf q} = {\bf v}_{\bf k-q} \ .
\label{extremum}
\end{equation}
Using this, the decay threshold boundary  is determined by  the system of
equations (\ref{energy_conserv}) and (\ref{extremum}).
One can distinguish several types of general solutions,
or decay channels, for such boundaries
 \cite{StatPhysII,Chernyshev06}:\\[0.5mm]
(i)\ The emission of two equivalent magnons with
equal momenta ${\bf q} = {\bf k-q} = \frac{1}{2}({\bf k+G})$, where
$\bf G$ is a reciprocal lattice vector. In this case  Eq.~(\ref{extremum})
is satisfied automatically while Eq.~(\ref{energy_conserv}) is reduced to
\begin{equation}
\varepsilon_{\bf k} =  2 \varepsilon_{({\bf k+G})/2} \ .
\label{equalQ}
\end{equation}
(ii)\ The emission of an acoustic magnon with ${\bf q\to 0}$.
In this case Eq.~(\ref{energy_conserv}) is fulfilled while  Eq.~(\ref{extremum})
is equivalent to
\begin{equation}
|{\bf v}_{\bf k}| = |{\bf v}_0|\ .
\label{equalV}
\end{equation}
(iii)\ The emission of an acoustic magnon with ${\bf q}\to {\bf Q}_i$ (for
magnets with several acoustic branches):
\begin{equation}
\varepsilon_{\bf k} = \varepsilon_{{\bf k}-{\bf Q}_i}\ .
\label{acousticQ}
\end{equation}
(iv)\ The emission of two finite-energy magnons with different energies but the same velocities.
No simplification occurs in this case.

The resultant decay region is given by the union of all subregions
enclosed by the boundaries obtained by considering all the  decay channels in  (i)--(iv).
Some of the calculated surfaces define the decay boundary
while the rest correspond to saddle-points inside the decay
region. Figure~\ref{regions} shows an example of the decay threshold boundaries
for the cubic Heisenberg antiferromagnet in applied magnetic field,
the problem discussed in more detail in Sec.~\ref{sec:field_decays}.
Magnetic fields are given in units of the saturation field $H_s$
and below $H^*\approx 0.76H_s$ decays are strictly forbidden.
For $H>H^*$ the decay region grows out of the M$_\pi$ point,
filling out the whole Brillouin zone at $H=H_s$.

\subsection{Decay of acoustic excitations}
\label{sec:kinematics_acoustic}

For a sufficiently complicated dispersion law $\varepsilon_{\bf k}$
the decay threshold boundary has to be found numerically
by solving Eq.~(\ref{energy_conserv}) and Eqs.~(\ref{extremum})--(\ref{acousticQ}).
However, it is often possible to draw conclusions about the presence of
decays by analyzing only the low-energy part of the spectrum.
Such an approach is frequently used in theoretical studies of phonon decays
\cite{Kosevich} and we briefly review
the corresponding arguments, which will also be relevant for the subsequent treatment
of magnon decays.

\bigskip\medskip\noindent
\subsubsection{Single acoustic branch}

In the decay process of a long-wavelength acoustic excitation
the incident and the two emitted quasiparticles have the same velocity
$c$. The linear approximation for the excitation energy,
$\varepsilon_{\bf k}= ck$, does not provide sufficient information on
the possibility of decays
and one has to include a nonlinear correction to the linear dispersion:
\begin{equation}
\varepsilon_{\bf k} \approx ck + \alpha k^3 \ .
\label{acoustic}
\end{equation}
For a weak nonlinearity  $\alpha k^3\ll ck$,  the momenta of decaying and
emitted quasiparticles are nearly parallel such that
\begin{equation}
|{\bf k}-{\bf q}| \approx k-q + \frac{kq\varphi^2}{2(k-q)} \ ,
\end{equation}
where $\varphi$ is a small angle between $\bf q$ and $\bf k$.
The energy conservation  (\ref{energy_conserv}) within the same approximation is
\begin{equation}
3\alpha kq(k-q) = \frac{ckq\varphi^2}{2(k-q)} \ .
\end{equation}
The nontrivial solutions $q, \varphi\neq 0$ exist only for the positive sign of
the cubic nonlinearity  $\alpha$. While the asymptotic form (\ref{acoustic}) is
relevant to many physical examples, an even more general condition can be
formulated to include other cases of gapless energy spectra. If the low-energy
part of the spectrum is a concave function of the momentum
$\partial^2\varepsilon/\partial k^2<0$, Fig.~\ref{fig:acoustic}(a), magnons
remain stable. On the other hand, for the upward curvature
$\partial^2\varepsilon/\partial k^2>0$, Fig.~\ref{fig:acoustic}(b), the low-energy
excitations are unstable with respect to spontaneous decays.

Note that according to the above criterion two-magnon decays are kinematically
allowed in Heisenberg ferromagnets. Indeed, the energy conservation equation
(\ref{energy_conserv}) can be satisfied for quasiparticles with a ferromagnetic dispersion
law $\varepsilon_{\bf k}\propto k^2$.
However, as discussed in Sec.~II, the decay vertices $V^{(1)}_{3}$ and $V^{(1)}_{4}$
vanish exactly in the Heisenberg ferromagnet and magnons remain well defined.
Nevertheless, decays do exist in anisotropic planar ferromagnets, for which  the magnon
number conservation does not hold anymore and $V^{(1)}_{4}$ is nonzero
\cite{Villain74,Stephanovich11}.

\subsubsection{Several acoustic branches}

\begin{figure}[t]
\includegraphics[width=0.99\columnwidth]{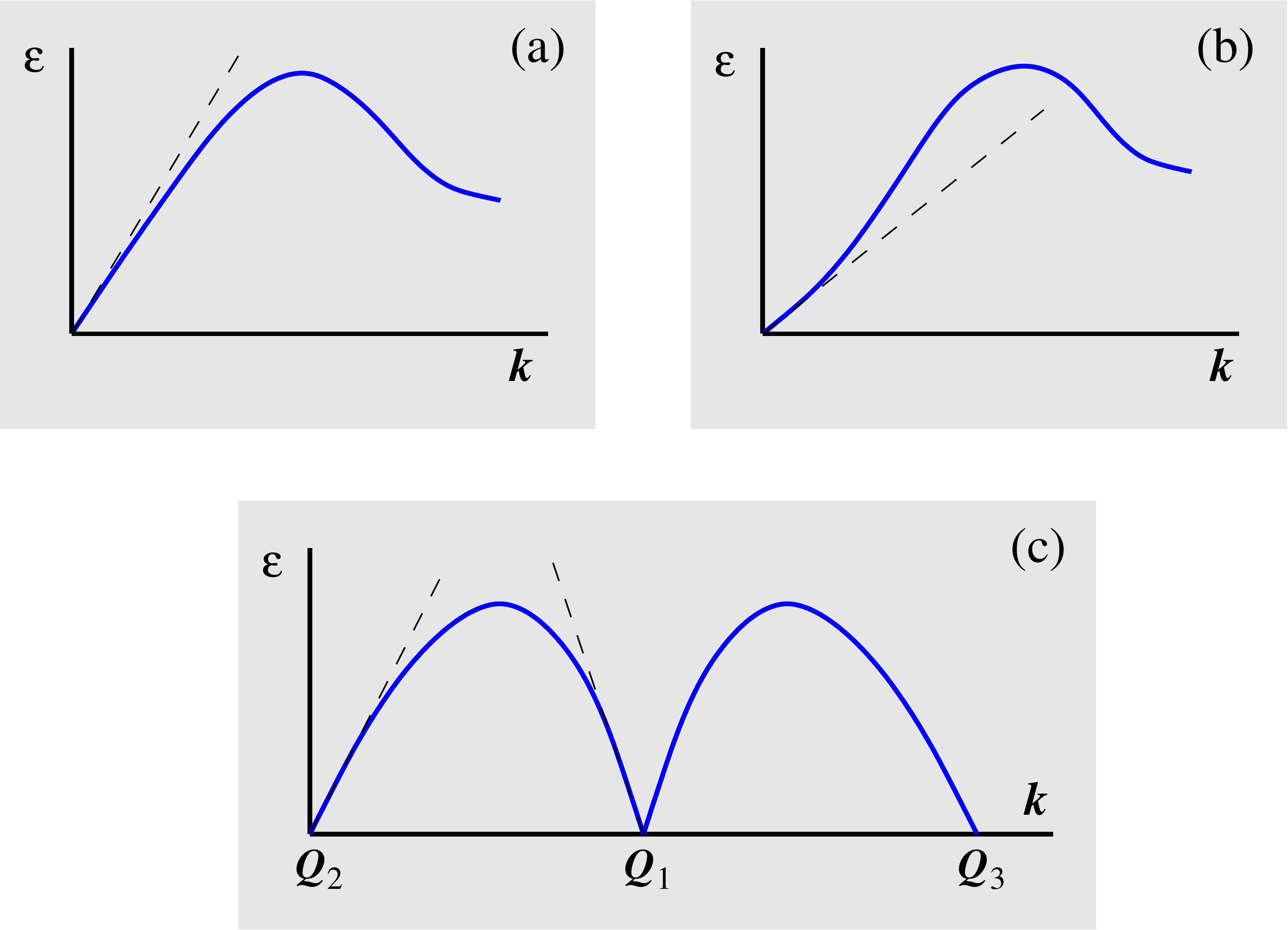}
\caption{(color online). Sketches of the quasiparticle spectra $\varepsilon_{\bf k}$
with (a) negative and (b) positive curvature of the acoustic mode,
respectively. (c) Spectrum with several acoustic branches.
}
\label{fig:acoustic}
\end{figure}

Lattice vibrations in crystals have three acoustic branches with different
sound velocities corresponding to one longitudinal and two transverse phonons.
Heisenberg antiferromagnets with a helical or spiral magnetic structure also
have three Goldstone modes, Fig.~\ref{fig:acoustic}(c),  at ${\bf k=0}$ and
$\pm\bf Q$, where $\bf Q$ is the ordering wave vector; see Sec.~\ref{sec:helix}
for more details.
Energy conservation for the decay of a fast quasiparticle into two
slow quasiparticles can be satisfied already in the linear approximation
for the energy
\begin{equation}
c_1 k = c_2 q + c_3 |{\bf k-q}|\ , \qquad \textrm{for}\quad c_1>c_2,c_3 \, ,
\end{equation}
where all momenta are measured relative to the corresponding Goldstone
${\bf Q}$ points. Thus, if this  decay channel is compatible with momentum
conservation, {\it i.e.}, ${\bf Q}_1 = {\bf Q}_2 + {\bf Q}_3$, then the fast
excitation can decay regardless of the curvature of the corresponding acoustic branch.

\subsection{Decay singularities}
\label{sec:kinematics_singular}

We now demonstrate the fact that the decay threshold boundary should
generally correspond to a nonanalyticity in the particle's spectrum because
of its  coupling to the two-particle continuum. Consider the decay
self-energy $\Sigma_a({\bf k},\omega)$ from Eq.~(\ref{SelfE3}) in the vicinity
of the threshold boundary for two-particle decays. Generally the decay
vertex has no additional smallness in terms of the momenta of participating
quasiparticles, {\it i.e.}, $V^{(1)}_{3} = {\cal O}(J)$. Focusing on the 2D case
and expanding Eq.~(\ref{energy_conserv}) in small $\Delta {\bf k} = {\bf k}-{\bf k}^*$
and $\Delta {\bf q} = {\bf q}-{\bf q}^*$, where ${\bf k}^*$ is the crossing
point of the single-magnon branch and the bottom of the continuum and ${\bf q}^*$
is the minimum point of $E_{2}({\bf k}^*,{\bf q})$, energy conservation gives
\begin{equation}
\varepsilon_{\bf k} -
\varepsilon_{\bf q} - \varepsilon_{\bf k-q} \approx
({\bf v}_1 - {\bf v}_2)\cdot \Delta {\bf k} -
\frac{\Delta q_x^2}{a^2} - \frac{\Delta q_y^2}{b^2} = 0\ ,
\label{conserv_expan}
\end{equation}
where ${\bf v}_1$ and ${\bf v}_2$ are the velocities of the initial and final
magnons and $a$ and $b$ are constants. Next, the singular part of the
self-energy is given by
\begin{equation}
\Sigma({\bf k},\varepsilon_{\bf k})  \propto
\int \frac{d^2q} {(v_1 -v_2)\Delta k - q_x^2/a^2 - q_y^2/b^2 + i0}\, ,
\label{sigmaB}
\end{equation}
and a straightforward integration in Eq.~(\ref{sigmaB}) yields a logarithmic
singularity in the spectrum \cite{Zhitomirsky99}
\begin{equation}
\textrm{Re}\Sigma({\bf k},\varepsilon_{\bf k})\simeq
\ln\frac{\Lambda}{|\Delta k|}\,, \
\Gamma_{\bf k} =-\textrm{Im}\Sigma({\bf k},\varepsilon_{\bf k})\simeq
\Theta(\Delta k) \,,
\label{logB}
\end{equation}
where $\Theta(x)$ is the step function. The imaginary part of
$\Sigma({\bf k},\varepsilon_{\bf k})$ in  Eq.~(\ref{Gamma2_k}) is directly related
to the two-particle density of states. Therefore, in two dimensions it is natural
to have a jump in $\Gamma_{\bf k}$ upon entering the continuum, as obtained in
Eq.~(\ref{logB}). In effect this demonstrates that the van Hove singularity in the
continuum's density of states gets transferred directly onto the quasiparticle
spectrum via the three-particle coupling.

Consider now the other type of 2D van Hove singularity, a saddle-point inside the
continuum. Expansion of the energy conservation in the vicinity of the saddle point yields
the same  form as in Eq.~(\ref{conserv_expan}) with a change of sign in front of either
$\Delta q_x^2$ or $\Delta q_y^2$. Therefore, the singular part of the self-energy is
\begin{equation}
\Sigma({\bf k},\varepsilon_{\bf k})  \propto
\int \frac{d^2q} {(v_1 -v_2)\Delta k - q_x^2/a^2 + q_y^2/b^2 + i0}\
\label{sigmaS}
\end{equation}
and the logarithmic singularity appears now in the imaginary part:
\begin{equation}
\textrm{Re}\Sigma({\bf k},\varepsilon_{\bf k}) \simeq
\textrm{sign}(\Delta k)\, ,
\ \ \ \
\Gamma_{\bf k} \simeq
\ln\frac{\Lambda}{|\Delta k|}\, ,
\label{logS}
\end{equation}
again in agreement with the relation between the 2D two-magnon DoS and $\Gamma_{\bf k}$.

Geometric consideration of the decay surface near singularities \cite{Chernyshev06}
offers a useful alternative perspective: while crossing of the decay threshold boundary
obviously corresponds to nucleation of the decay surface, the saddle-point singularity
corresponds to splitting of the decay surface into two disjoint pieces. Altogether, both
types of singularities (\ref{logB}) and (\ref{logS}) result from topological transitions
of the decay surface with a change in the number of connected sheets.

In 3D the decay self-energy generally exhibits a less singular but still nonanalytic behavior.
The logarithmic peak at the decay threshold boundary is replaced by a continuous square-root
behavior $\simeq \sqrt{|\Delta k|}$ for both $\textrm{Re}\Sigma({\bf k},\varepsilon_{\bf k})$
and $\Gamma_{\bf k}$. A remarkable exception  is the threshold to the two-roton decay in the
spectrum of superfluid $^4$He. The dispersion of roton excitations has a minimum on a sphere
in the momentum space, leading to the same logarithmic divergence of the self-energy as in
Eq.~(\ref{logB}) at the energy twice the roton gap  \cite{StatPhysII}. This strong anomaly produces
the ``endpoint'' in the dispersion curve $\varepsilon_{\bf k}$ of the superfluid $^4$He predicted
by \textcite{Pitaevskii59}. Such a termination point was later observed in inelastic
neutron-scattering experiments \cite{Graf74,Glyde98}.

The analysis of this section can be straightforwardly extended to the three-particle
decay self-energy $\Sigma_c({\bf k},\omega)$ given by Eq.~(\ref{SelfE4}). The general outcome
is similar to the effects of higher dimensions: singularities are replaced by continuous
albeit nonanalytic behavior. Overall, the two-magnon decays play a more significant
role than the higher-order decay processes. Decays and associated singularities are also
enhanced by the low dimensionality of a magnetic system.

\section{Field-induced decays}
\label{sec:field_decays}

We devote this section to spontaneous decays in the square-lattice Heisenberg
antiferromagnet in strong external field. In zero field the ground state of
this model is a two-sublattice N\'eel structure, which, due to its collinearity,
prohibits cubic anharmonicities and two-magnon decays;
see Sec.~\ref{sec:anharmonic}.B. The three-magnon and all other
$n$-magnon decays are forbidden by energy conservation, as in
the cubic-lattice model studied by \textcite{Harris71} and in agreement with
the arguments of Sec.~\ref{sec:kinematics}. Thus, in zero field and at $T=0$
magnons have an infinite lifetime. In the opposite limit, when spins are fully
polarized by external field, the ground state is equivalent to that of
a ferromagnet with all particle-nonconserving terms strictly forbidden so
that the magnons are well defined again. It comes as a surprise that
there exists a regime between these two limits where magnons become heavily
damped throughout the Brillouin zone.

Next we discuss general arguments for the existence of the
field-induced decays in a broad class of antiferromagnets
and provide details of the spin-wave approach to the  square-lattice
antiferromagnet in a field. Focusing mostly on the $S=1/2$ case,  we
demonstrate that magnons become damped and even overdamped in most of
the Brillouin zone for a range of fields near the saturation field
\cite{Zhitomirsky99}. Confirmation of these results from numerical
\cite{Olav08} and experimental studies is highlighted. We also illustrate
the occurrence of the threshold and saddle-point singularities discussed
in Sec.~\ref{sec:kinematics}.C in the perturbative treatment of the
magnon spectrum and offer two approaches to regularize them self-consistently.
Detailed results of one such self-consistent regularization method \cite{Mourigal10}
are presented and an example of the higher-dimensional extension of the
square-lattice case is touched upon \cite{Fuhrman12}.

\begin{figure}[t]
\centering
\includegraphics[width=0.8\columnwidth]{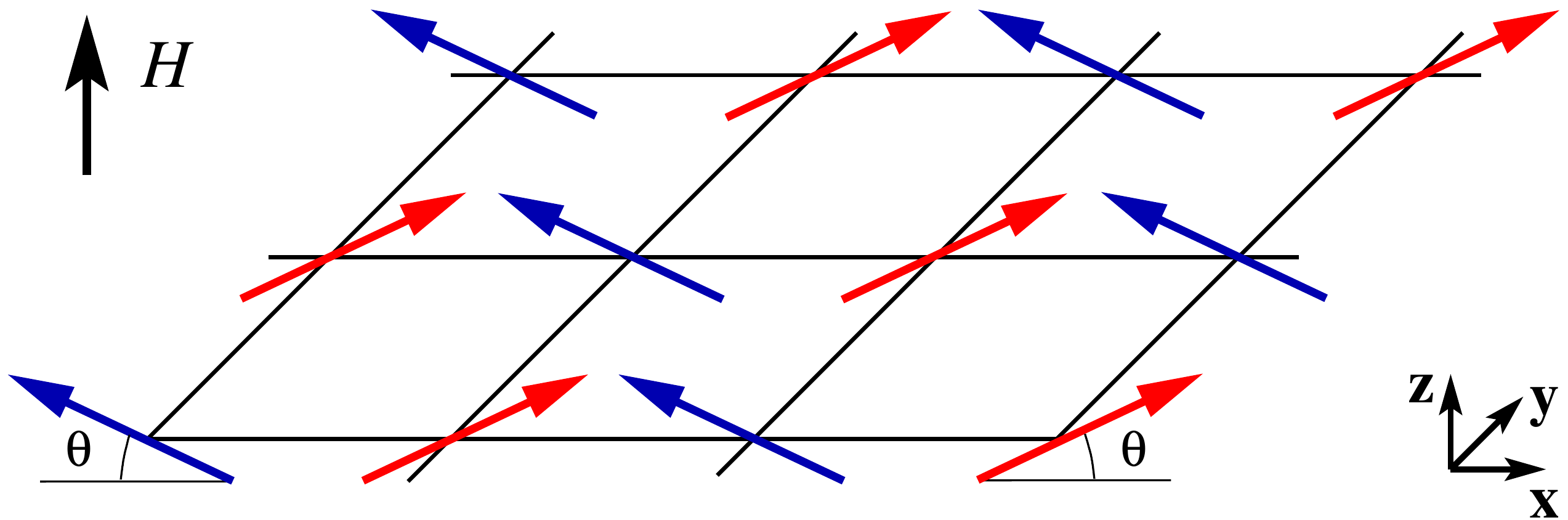}
\caption{(color online).\ \
Field-induced canted spin structure in the square-lattice Heisenberg antiferromagnet.
}
\label{fig:canted_structure}
\end{figure}

\subsection{General discussion}

The evolution of an ordered magnetic structure of the two-sublattice
antiferromagnet in an external field is relatively trivial.
Spins orient themselves transverse to the field direction and
tilt gradually towards its direction until they become fully aligned
at $H\geq H_s$, where $H_s$ is referred to as the
saturation field, see Fig.~\ref{fig:canted_structure}.
In the absence of anisotropy and frustration there is no other phase transition
in the whole range from $H=0$ to $H=H_s$.
As first argued by \textcite{Zhitomirsky99}, it is
the dynamical properties of quantum antiferromagnets that
undergo an unexpectedly dramatic transformation on the way
to full saturation. The phenomenon responsible
for this transformation is spontaneous magnon decay, which becomes
possible in sufficiently strong magnetic fields.

The field-induced magnon decays are generally present in a broad class
of quantum antiferromagnets. Indeed,
three-magnon interactions that couple one- and two-magnon states always exist in
canted antiferromagnetic structures created by an external field; see
Sec.~\ref{sec:anharmonic}. Usually, kinematic constraints prevent
spontaneous decays from taking place in zero or weak magnetic fields.
It is easy to see that from the acoustic branch of the magnon spectrum
(\ref{acoustic}), which retains the concave shape ($\alpha<0$)
in small fields and thus is stable with respect to spontaneous decays according
to Sec.~\ref{sec:kinematics}.B. However, in a strong enough magnetic field
the convexity of the acoustic spectrum must change. This is because at
the saturation field, $H=H_s$,  the magnon velocity vanishes and the dispersion
at low energies is that of a ferromagnet, $\varepsilon_{\bf k} \propto k^2$,
which has an upward curvature. By continuity, the spectrum has to preserve
its positive curvature for a certain range of magnetic fields $H^*<H<H_s$,
where the threshold field $H^*$ corresponds to $\alpha=0$. This  implies
that the two-particle decays are present at $H>H^*$.

\subsection{Square-lattice antiferromagnet}
\label{sec:safm_SWT}

The Heisenberg model on a square lattice, a toy model in the early days of
the theory of aniferromagnets \cite{Anderson52,Kubo52},  became a
paradigmatic model in the late 1980s due to its direct relevance to the
parental materials of the high-$T_c$ cuprates \cite{Chakravarty89,Manousakis91}.
It is also viewed as one of the prototypical strongly correlated
models in condensed matter physics, used for benchmarking various theoretical
methods against the more precise numerical techniques \cite{Sandvik97,White07}.

The standard spin-wave theory provides an accurate description of the static
and dynamical properties of the square-lattice antiferromagnet in zero field
even for $S=1/2$ \cite{Hamer92,Weihong93,Igarashi92,Igarashi05}. In the words
of \textcite{Sandvik97}: ``The presently most accurate calculations [for
the $S=1/2$ case]... indicate that the true values of the ground-state
parameters deviate from their $1/S^2$ spin-wave values by only 1-2\% or less.''
The spin-wave calculations also offer a fine fit to the overall magnon dispersion
\cite{Weihong93,Igarashi05,Syromyatnikov10} with the only deviation from the numerical
results along the $(\pi,0)$-$(0,\pi)$ path \cite{Zheng05,Sandvik01}.

Heisenberg models in strong fields have received somewhat less attention until
recently [see, however, \textcite{deJongh01} and \textcite{deGroot86}], primarily because
the majority of available materials had exchange constants much larger than
the strength of available magnetic fields. The constraints are even more
stringent on neutron-scattering studies in the field, where the practical
limit is currently at about 14T. Nevertheless, recent developments in
the synthesis of molecular based antiferromagnets with moderate strength of
exchange coupling between spins \cite{Woodward02,Lancaster07,Coomer07,Xiao09}
have opened the high-magnetic-field regime  to experimental investigations for
a number of layered square-lattice materials \cite{Tsyrulin09,Tsyrulin10}.
Furthermore,  new field-induced dynamical effects can be present in
antiferromagnets with other lattice geometries \cite{Coldea01}.

Another class of antiferromagnets directly relevant to our discussion
incorporates quantum spin systems with singlet ground states and gapped
triplet excitations. They are often called BEC magnets because the
field-induced ordering in them can be described in terms of the Bose-Einstein
condensation (BEC) of triplet excitations. In recent years these quantum
spin-gap magnets have been intensively studied both experimentally
\cite{Nikuni00,Jaime04,Regnault06} and theoretically
\cite{Affleck91,Affleck92,Mila98,Giamarchi99} and continue to attract
attention \cite{Giamarchi08}. We point out that
the mechanism of spontaneous magnon decays discussed here equally applies
to  the vicinity of the triplet condensation field $H_c$ in these magnets
because of the duality between $H_c$ and the saturation field $H_s$
\cite{Mila98}. This is highly advantageous because in many BEC magnets
the lower condensation field is readily accessible, which makes them
prime candidates for observing magnon decays directly in inelastic
neutron-scattering experiments. It can  be argued that the recently observed
suppression of thermal conductivity in the vicinity of critical fields in
one such material \cite{Kohama2011} is due to magnon decays.

Extension of the spin-wave theory to finite magnetic fields was developed by
\textcite{Osano82}, \textcite{Zhitomirsky98}, \textcite{Zhitomirsky99},
\textcite{Chernyshev09b}, and \textcite{Mourigal10}.
For alternative approaches applicable in the vicinity of the saturation field,
see also \textcite{Batyev84}, \textcite{Gluzman93}, \textcite{Kreisel08},
and \textcite{Syromyatnikov09}.
Close agreement of the spin-wave  calculations with exact diagonalization
and quantum Monte Carlo (QMC)
results for the ground-state properties in the whole range of fields
$0<H<H_s$ was established by \textcite{Luscher09}.
The same work has also confirmed the earlier QMC results by \textcite{Olav08} that
demonstrated clear signatures of magnon decays in the high-field regime.
We provide more details on that in the following.

\subsubsection{Model and formalism}
\label{sec:formalities}

We begin with the Heisenberg Hamiltonian of nearest-neighbor
spins on a square lattice in a magnetic field directed
along $z_0$ axis in the laboratory reference frame,
\begin{equation}
\hat{\cal H} = J \sum_{\langle ij \rangle}
\mathbf{S}_i\cdot\mathbf{S}_j -   H \sum_{i}  S_i^{z_0}  \ .
\label{H_safm}
\end{equation}
Here, $J$ is the nearest-neighbor  coupling and $H$ is the field in units
of $g\mu_B$. With the details of the technical approach explicated by
\textcite{Zhitomirsky98} and \textcite{Mourigal10}, we summarize here the key
steps of the spin-wave theory approach to this problem.

First we align the local spin quantization axis on each site in the direction given
by the canted spin configuration shown in Fig.~\ref{fig:canted_structure}. The corresponding
transformation of the spin components from the laboratory frame $(x_0,y_0,z_0)$
to the local frame is
\begin{eqnarray}
 && S_i^{x_0} =  S_i^x \sin\theta + S_i^z e^{i {\bf Q}\cdot {\bf r}_i} \cos\theta \ ,
     \quad S_i^{y_0}=S_i^{y} \ ,
 \nonumber \\
 && S_i^{z_0} = -S_i^x e^{i{\bf Q}\cdot{\bf r}_i} \cos\theta + S_i^z \sin\theta \ ,
\label{local_frame}	
\end{eqnarray}
where $\theta$ is the canting angle to be defined from the energy minimization
and ${\bf Q}=(\pi,\pi)$ is the ordering wave vector of the two-sublattice N\'eel
structure. Next the standard Holstein-Primakoff transformation bosonizes the spin
operators $S^z_i = S - a^\dagger_ia_i$, $S^-_i = a^\dagger_i (2S-a^\dagger_ia_i)^{1/2}$,
with $S^\pm_i = S_i^x \pm iS_i^y$. Expanding square-roots, one obtains the bosonic Hamiltonian
as a sum,
\begin{equation}
\hat{\cal H}_{SW} = \hat{\cal H}_0 + \hat{\cal H}_1 + \hat{\cal H}_2+
\hat{\cal H}_3 + \hat{\cal H}_4 + \dots,
\label{HSW}
\end{equation}
each term being of the $n$th order in bosonic operators $a^\dagger_i$ and $a_i$
and carrying an explicit factor $S^{2-n/2}$. This form provides the basis for
the $1/S$ expansion. Important consequences of the noncollinear spin structure
are the terms $S_j^zS_i^{x(y)}$, which describe coupling between transverse
$S^{x(y)}$ and longitudinal $S^z$ oscillations and yield cubic anharmonicities
($\hat{\cal H}_{3}$) in the bosonic Hamiltonian.

Minimization of the classical energy $\hat{\cal H}_0$  in Eq.~(\ref{HSW}) fixes
the canting angle of the classical spin structure to $\sin\theta= H/H_s$,
where the saturation field is $H_s = 8JS$. This procedure also cancels the
$\hat{\cal H}_1$ term that is linear in $a^\dagger_i$ and $a_i$. After
subsequent Fourier transformation, the harmonic part of the Hamiltonian
($\hat{\cal H}_2$) is diagonalized by the Bogolyubov transformation,
$a_{\bf k} = u_{\bf k} b_{\bf k} + v_{\bf k} b^\dagger_{-\bf k}$, yielding
\begin{eqnarray}
&&\hat{\cal H}_{SW} = \sum_{\bf k} \tilde{\varepsilon}_{\bf k}
b_{\bf k }^{\dagger} b_{\bf k}^{\phantom{\dagger}} \label{eq:cubic} \\
&&\phantom{\hat{\cal H} =}
 + \frac{1}{2} \sum_{\bf k, q}  V_{3}^{(1)}({\bf k, q})
 \bigl( b_{\bf k-q+Q}^{\dagger} b_{\bf q }^{\dagger}
 b_{\bf k }^{\phantom{\dagger}} + \rm{h.c.} \bigr) +\ldots,
\nonumber
\end{eqnarray}
where
$\tilde{\varepsilon}_{\bf k} = \varepsilon_{\bf k} + \delta\varepsilon_{\bf k}$ with
$\varepsilon_{\bf k}$ being the magnon dispersion given by the linear spin-wave theory and
$\delta\varepsilon_{\bf k}$ is  from the $1/S$ corrections due to angle renormalization
and Hartree-Fock decoupling of cubic and quartic terms in Eq.~(\ref{HSW}), respectively.
The ellipsis stands for the classical energy, the three-boson source term $V_{3}^{(2)}$ as
in Eq.~(\ref{V3}), and   the higher-order terms in  the $1/S$ expansion.
Although some of these terms do contribute to the subsequent results, this abbreviated form
constitutes the essential part of the Hamiltonian and is sufficient for our discussion.
For the explicit expressions of  $\delta\varepsilon_{\bf k}$  and
vertices $V_{3}^{(1)}$ and $V_{3}^{(2)}$, see \textcite{Mourigal10}.

\begin{figure}[t]
\begin{center}
\includegraphics[width=0.65\columnwidth]{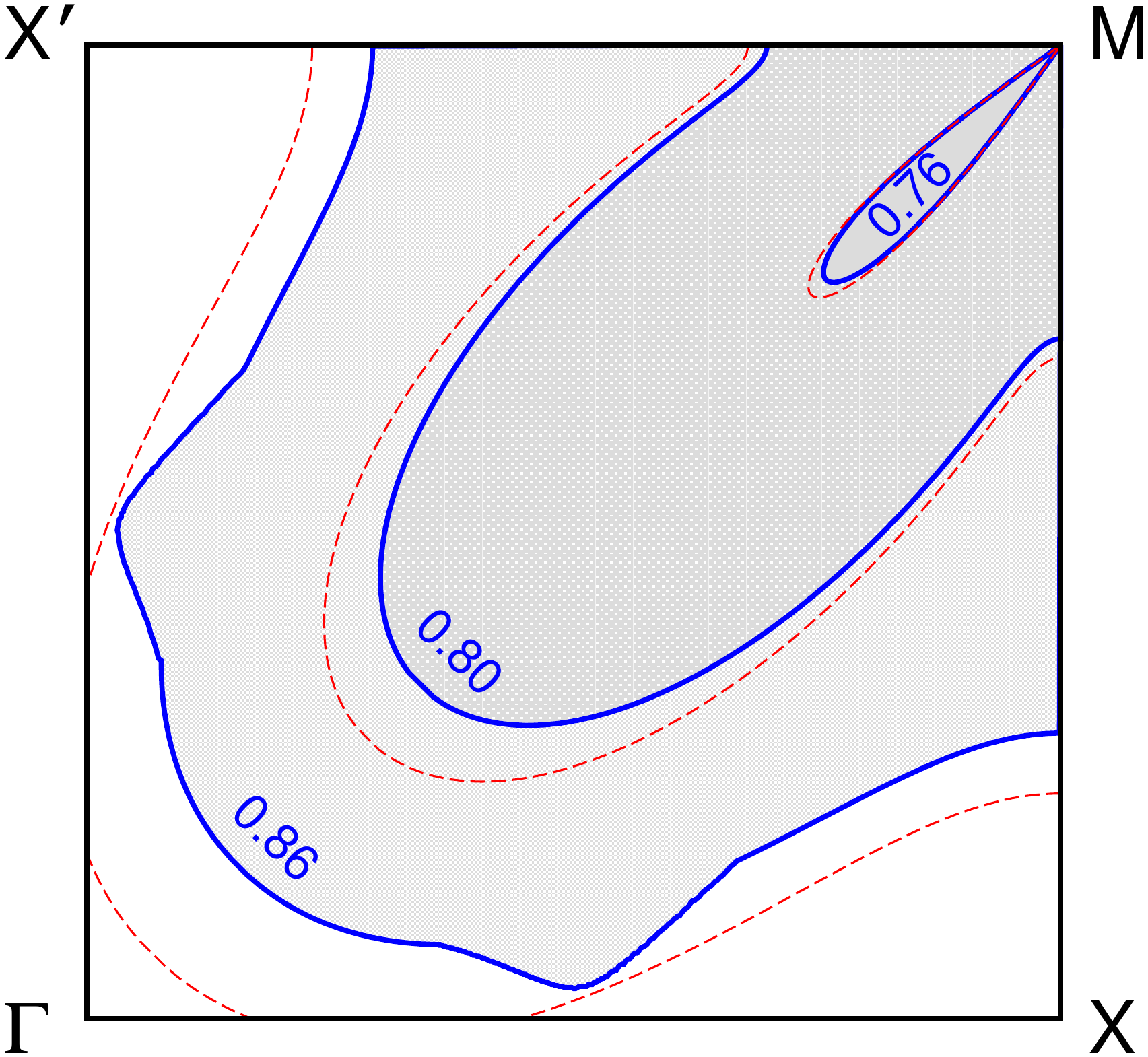}
\end{center}
\caption{(color online).\ \
Decay regions in one-fourth of the Brillouin zone
for representative values of $H/H_s$ indicated by numbers.
Solid and dashed lines are threshold boundaries for two- and three-magnon
decays, respectively.
}
\label{fig:decayregion_safm}
\end{figure}

\subsubsection{Decay regions}
\label{sec:sq_kinematics}

One can already gain detailed insight into the decay conditions in the manner
discussed in Sec.~\ref{sec:kinematics} by analyzing the magnon spectrum
in the harmonic approximation, which is given by
\begin{equation}
\varepsilon_{\bf k} = 4JS\sqrt{(1+\gamma_{\bf k})(1-\cos 2 \theta \gamma_{\bf k})}\ ,
\label{Ek_safm}
\end{equation}
where $\gamma_{\bf k} ={\textstyle \frac{1}{2}} (\cos k_x + \cos k_y)$.
The acoustic mode near ${\bf Q}$
follows the asymptotic form (\ref{acoustic})   $\varepsilon_{\bf Q+k} \approx c k + \alpha k^3$ with
\begin{eqnarray}
c = 2\sqrt{2}JS\cos\theta\,, \
\alpha = \frac{c}{16}
\left(\tan^2{\theta}-\frac{9+\cos{4\varphi}}{6} \right),
\label{E_acoust}
\end{eqnarray}
where $\varphi$ is the azimuthal angle of ${\bf k}=(k_x,k_y)$.
The curvature of the spectrum  $\alpha$ changes its sign
for ${\bf k}$ along the diagonal ($\varphi = \pi/4$) at the decay threshold field
\begin{equation}
H^* = \frac{2}{\sqrt{7}} \ H_{s} \approx 0.7559 \,H_s\ .
\label{H*}
\end{equation}
Note that this expression remains valid for the cubic- and  for the layered
square-lattice antiferromagnet with arbitrary antiferromagnetic interlayer coupling.

Staggered canting of spins in Eq.~(\ref{local_frame}) ``shifts'' the momentum in
the two-magnon decay  condition in Eq.~(\ref{energy_conserv}), which now reads as
$\varepsilon_{\bf k} = \varepsilon_{\bf q} + \varepsilon_{\bf k-q+Q}$.
Using the Bose-condensate perspective of Sec.~\ref{sec:3mag_sec}, such a change
is natural as the magnon going into the condensate now carries the momentum
$\bf Q$; see Eq.~(\ref{condensate}). Apart from that the kinematic analysis of
Sec.~\ref{sec:kinematics} remains the same. Numerical results for the decay
region at $H>H^*$ are shown in Fig.~\ref{fig:decayregion_safm}. Up to
$H\approx 0.84H_s$ the boundary of the decay region is determined  entirely
by the decays into a pair of magnons with equal momenta; see
Sec.~\ref{sec:kinematics_threshold}, type (i) in Eq.~(\ref{equalQ}). At fields
higher  than $0.85 H_s$ the decay channel with emission of an acoustic magnon,
type (ii) in Eq.~(\ref{equalV}), creates protrusions leading to more complicated
shapes of the decay region. Above $0.9H_s$ the decay region spreads almost
over the entire Brillouin zone, similarly to the 3D decay regions in Fig.~\ref{regions}.
In Fig.~\ref{fig:decayregion_safm}, we also show boundaries for the
three-magnon decays, which are less restrictive and cover larger areas
of the Brillouin zone than their two-magnon counterparts.

\subsection{Spectral function and self-consistent Born approximation}
\label{sec:Akw_1}

Information on the magnon  energy renormalization and
decay-induced lifetime due to  interactions can be obtained from the
single-particle spectral function
${A}({\bf k}, \omega)\!=\! -(1/\pi)\,{\rm Im}\,G({\bf k}, \omega)$.
Here $G({\bf k}, \omega)$ is the interacting Green's function,
\begin{equation}
 G^{-1}({\bf k}, \omega)  = \omega - \tilde{\varepsilon}_{\bf k}
 - \Sigma_{a}({\bf k}, \omega) - \Sigma_{b}({\bf k}, \omega)
\label{eq:green}	
\end{equation}
where $\Sigma_{a, b}({\bf k}, \omega)$ are the decay and source self-energies
shown in Fig.~\ref{fig:diagrams},
with the ``one-loop'' expressions for them given in Eq.~(\ref{SelfE3}).
Importantly, the magnon spectral function is directly related to the
dynamical structure factor $S({\bf k},\omega)$ measured in neutron-scattering experiments;
see \textcite{Mourigal10} for the details of their connection.

\begin{figure}[t]
\centerline{
\includegraphics[width = 0.7\columnwidth]{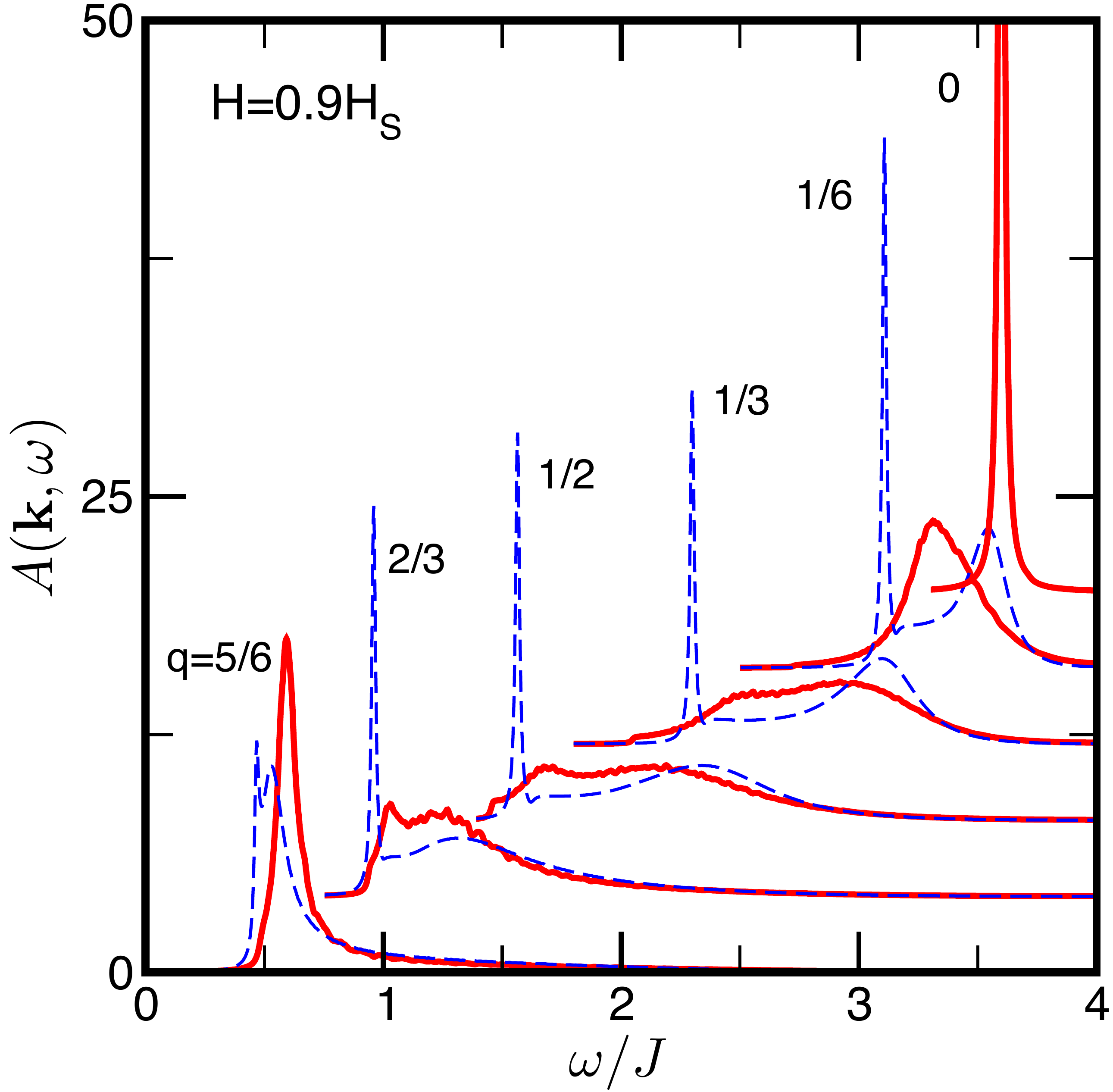}
}
\caption{(color online).\ \
Magnon spectral function for the spin-$\frac{1}{2}$ antiferromagnet
at $H = 0.9H_s$ for ${\bf k}=\pi (q,q)$.
Dashed and solid lines correspond to Born approximation and
restricted self-consistent Born approximation, respectively.
From  \textcite{Zhitomirsky99}.
}
\label{fig:Sxx1999}
\end{figure}

\begin{figure}[t]
\centerline{
\includegraphics[width=0.65\columnwidth]{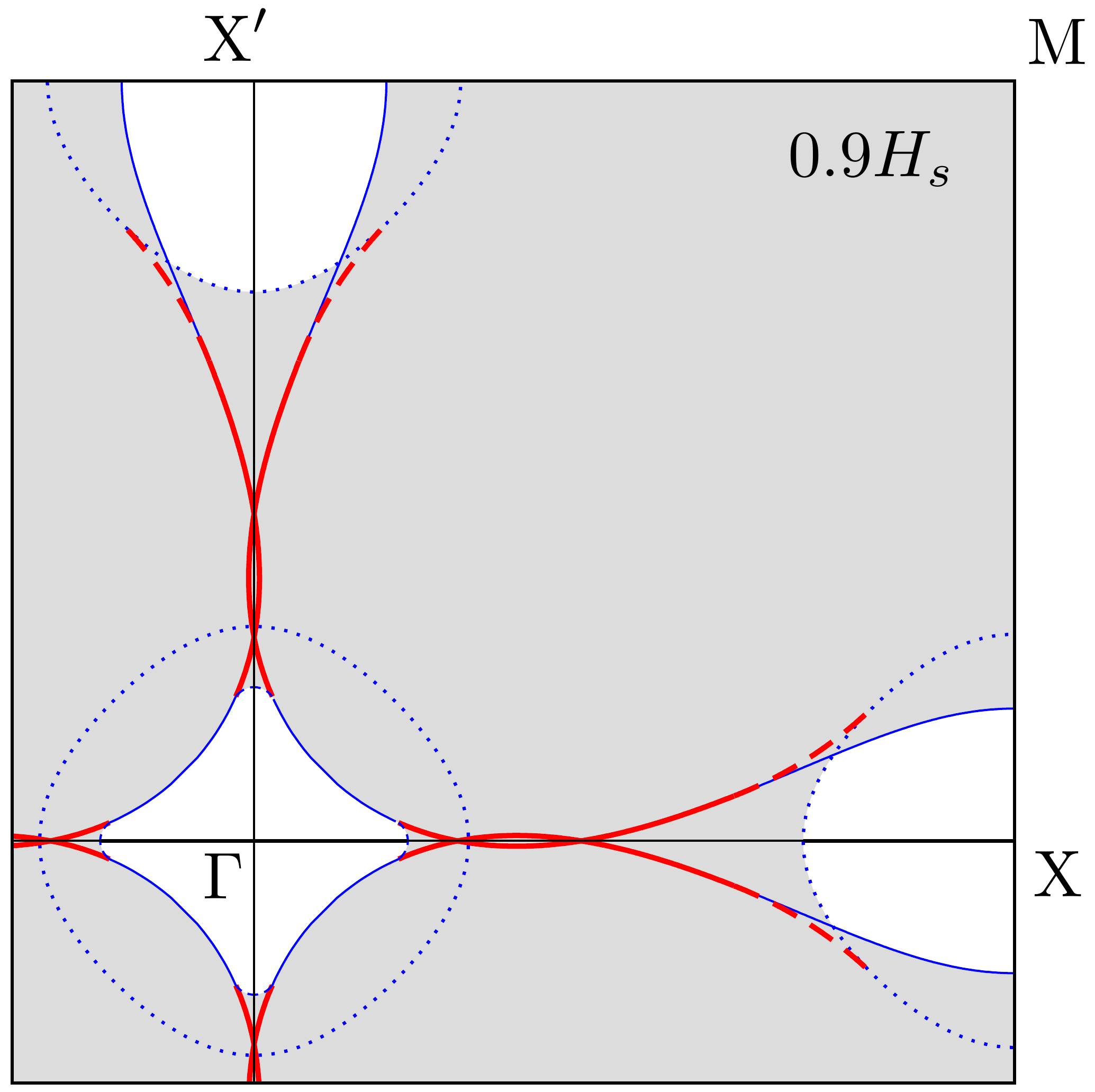}}
\caption{(color online).
Decay region and associated singularities for $H = 0.9H_s$. Solid, dashed, and
dotted lines denote the decay channels into a pair of magnons with equal
and different momenta, and emission of an acoustic magnon, respectively. Thick
portions of solid and dashed lines indicate the saddle-point
singularity in the continuum, while the thin parts  correspond to
local minima.
}
\label{fig:decayregion_09}
\end{figure}

The self-energies in Eq.~(\ref{eq:green}) originate from  coupling to the
two-magnon continuum and already in the lowest order  one  expects the spectral
function to exhibit an incoherent component due to two-magnon states in addition
to the quasiparticle peak that is broadened due to decays. The Born approximation
(BA) results for $A({\bf k},\omega)$
for several momenta along the $\Gamma M$ line at  $H=0.9H_s$
are shown in Fig.~\ref{fig:Sxx1999} by dashed lines.
While $A({\bf k},\omega)$ exhibits incoherent subbands,
the  peaks survive even for $\bf k$ deep inside the
decay region in Fig.~\ref{fig:decayregion_09}.
This is due to the lack of self-consistency in the lowest-order
BA, as  one- and two-magnon  states are treated with different accuracy.
Specifically, the magnon energy is renormalized downward by the real
part of $\Sigma_{a,b}({\bf k}, \omega)$  in Eq.~(\ref{eq:green}),
whereas the energy of the continuum is still calculated in the harmonic
approximation.
Taking into account  energy renormalization of the continuum
means ``dressing'' of the  inner lines of the decay diagram in Fig.~\ref{fig:diagrams}.
The main technical difficulty with such a self-consistent Born approximation (SCBA)
is that it opens
an unphysical gap for long-wavelength magnons in violation of the Goldstone theorem.

This difficulty notwithstanding, the restricted SCBA scheme was implemented
by \textcite{Zhitomirsky99}, which included partial renormalization
of the two-magnon continuum. It enforced  the acoustic form of the long-wavelength
magnon dispersion (\ref{acoustic}) and incorporated the self-consistent
renormalization of one inner magnon line in the decay diagram of Fig.~\ref{fig:diagrams}(a).
The results of the numerical solution of Dyson's equation in the restricted SCBA  are
shown in Fig.~\ref{fig:Sxx1999} by solid lines.
Not only are the  peaks that were artifacts of the non-self-consistent treatment
completely wiped out, but the overall values of  magnon damping
are large away from the ${\bf Q}$ and $\Gamma$ points,
the latter corresponding to the uniform precession mode which is stable.

\begin{figure}[b]
\centerline{
\includegraphics[width = 0.9\columnwidth]{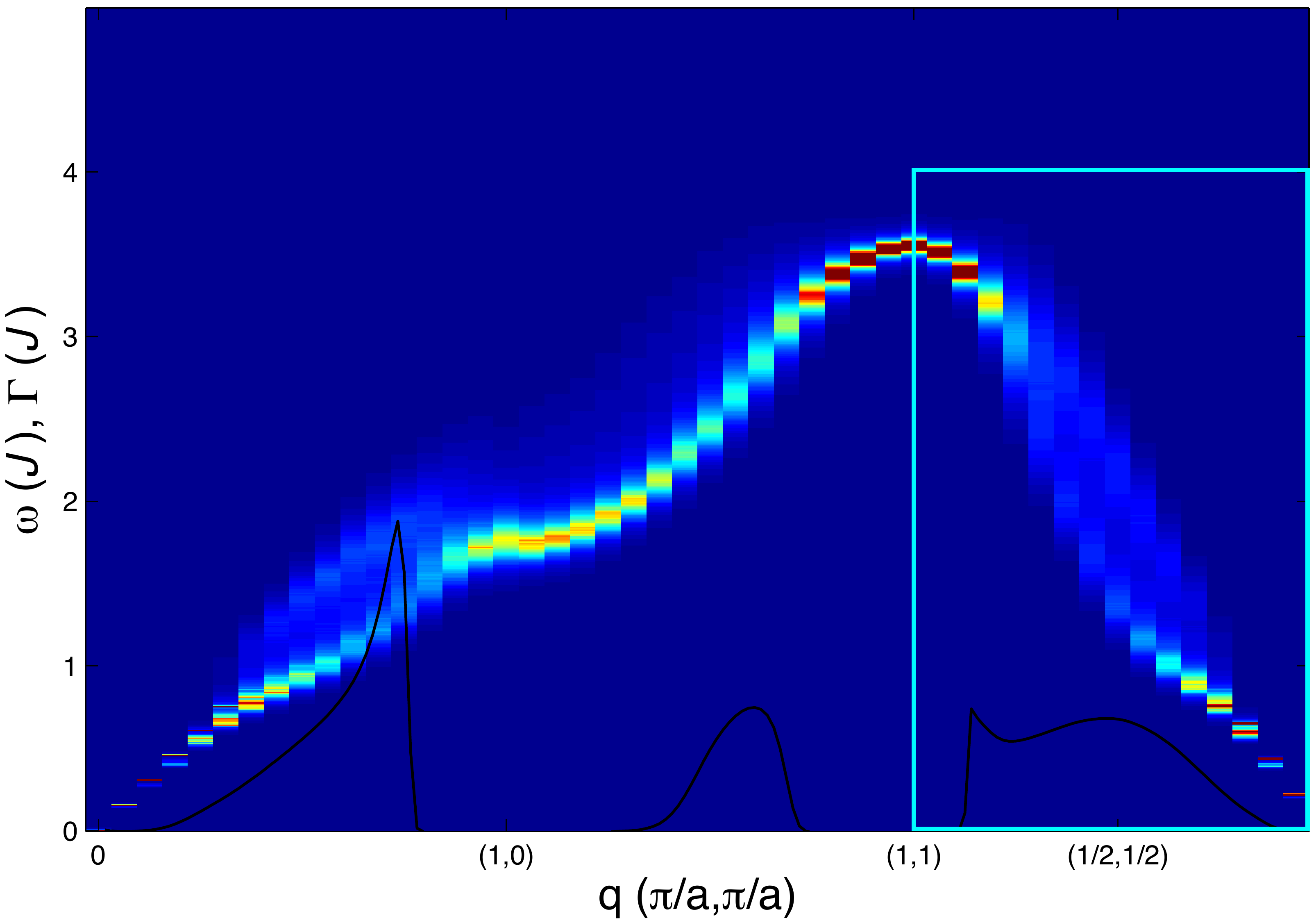}
}
\caption{(color online).\ \
QMC results for the $S^{zz}({\bf q},\omega)$ component of the dynamical structure
factor for the $S=1/2$ square-lattice antiferromagnet
at $H=0.875H_s$ along the $\Gamma X M\Gamma$ path.
Black lines are the magnon linewidths in the Born approximation \cite{Olav08}.
The box marks the ${\bf k}$-$\omega$ region
of Fig.~\ref{fig:Sxx1999} for comparison. From Olav Sylju\aa sen, 2008.
}
\label{fig:OlavSqw}
\end{figure}

These results received strong confirmation from the numerical QMC simulations
by \textcite{Olav08} (see Fig.~\ref{fig:OlavSqw}) and,
subsequently, from  the exact diagonalization (ED) study by \textcite{Luscher09}.
Both numerical works found significant modifications in the magnon
spectra above the threshold field: an explicit broadening of peaks in the
dynamical structure factor revealed by QMC data and a qualitative transformation in the
structure of energy levels and in distribution of spectral weights observed in ED.
Figure~\ref{fig:OlavSqw} shows the intensity plot of the $S^{zz}({\bf q},\omega)$
component of the dynamical structure factor from \textcite{Olav08}. Since
$S^{zz}({\bf q},\omega)$ is  proportional to the spectral function with a
shifted momentum $A({\bf q}-{\bf Q},\omega)$ \cite{Fuhrman12}, gapless and
uniform precession modes trade their places in Fig.~\ref{fig:OlavSqw}.

There are close qualitative and quantitative similarities between the
analytical  and numerical results of Figs.~\ref{fig:Sxx1999} and \ref{fig:OlavSqw}.
Magnon decays and spectrum broadening are strongly pronounced and follow  decay
regions suggested by  the kinematic analysis of Sec.~IV.B.
Sylju\aa sen also emphasized  the nontrivial redistribution of spectral weight
resulting in non-Lorentzian ``double-peak'' features in the dynamical structure factor, also
seen  in the  analytical results in Fig.~\ref{fig:Sxx1999}, albeit less pronounced.
In spite of the approximations of the analytical approach and
of the uncertainties in the QMC results due to numerical
interpolation from imaginary to real frequencies, such
a correspondence between  two very different methods is truly remarkable.

Following the original prediction of the field-induced spontaneous magnon decays,
there is an ongoing search for suitable spin-$\frac{1}{2}$ square-lattice materials
\cite{Lancaster07,Coomer07,Tsyrulin09} to investigate corresponding
effects. Recently, \textcite{Masuda10} found indications
of the field-induced decays in the inelastic neutron-scattering spectra
of the spin-$\frac{5}{2}$ layered square-lattice
antiferromagnet $\rm Ba_2MnGe_2O_7$.

\subsection{Decay singularities and SCBA}
\label{sec:safm_selfSWT}

While the preceding consideration demonstrates the spectacular role of
spontaneous decays in reshaping the magnon spectrum in the high-field regime
of the $S=1/2$ square-lattice antiferromagnet, several aspects of the bigger
picture remain unclear. In particular, what about the threshold and
saddle-point singularities discussed in Sec.~\ref{sec:kinematics}?  Are there
termination points in the magnon spectrum? How strong is the magnon damping in
more quasiclassical, $S\geq 1$, systems? These questions were addressed
recently by \textcite{Mourigal10}.

First we substantiate the generic analysis of the threshold singularities
provided in Sec.~\ref{sec:kinematics} by giving a specific example from the
square-lattice antiferromagnet. In Fig.~\ref{fig:decayregion_09} we show
the decay region for  $H = 0.9H_s$ together with the set of contours obtained
by the analysis of Sec.~\ref{sec:kinematics_threshold}. The solid, dashed,
and dotted lines are the singularity contours for the decay into magnons with
the same (\ref{equalQ}) and different momenta, and for the emission of an
acoustic magnon (\ref{equalV}), respectively. Line thickness indicates
a singularity that corresponds to a minimum (saddle point) of the continuum.
The same singularity contour may correspond to a minimum in one part of
the Brillouin zone  and to a saddle point in the other. This determines
whether the logarithmic singularity will occur in the real or imaginary
part of the self-energy, as discussed in Sec.~\ref{sec:kinematics_singular}.

\begin{figure}[b]
\centerline{
\includegraphics[width = 0.85\columnwidth]{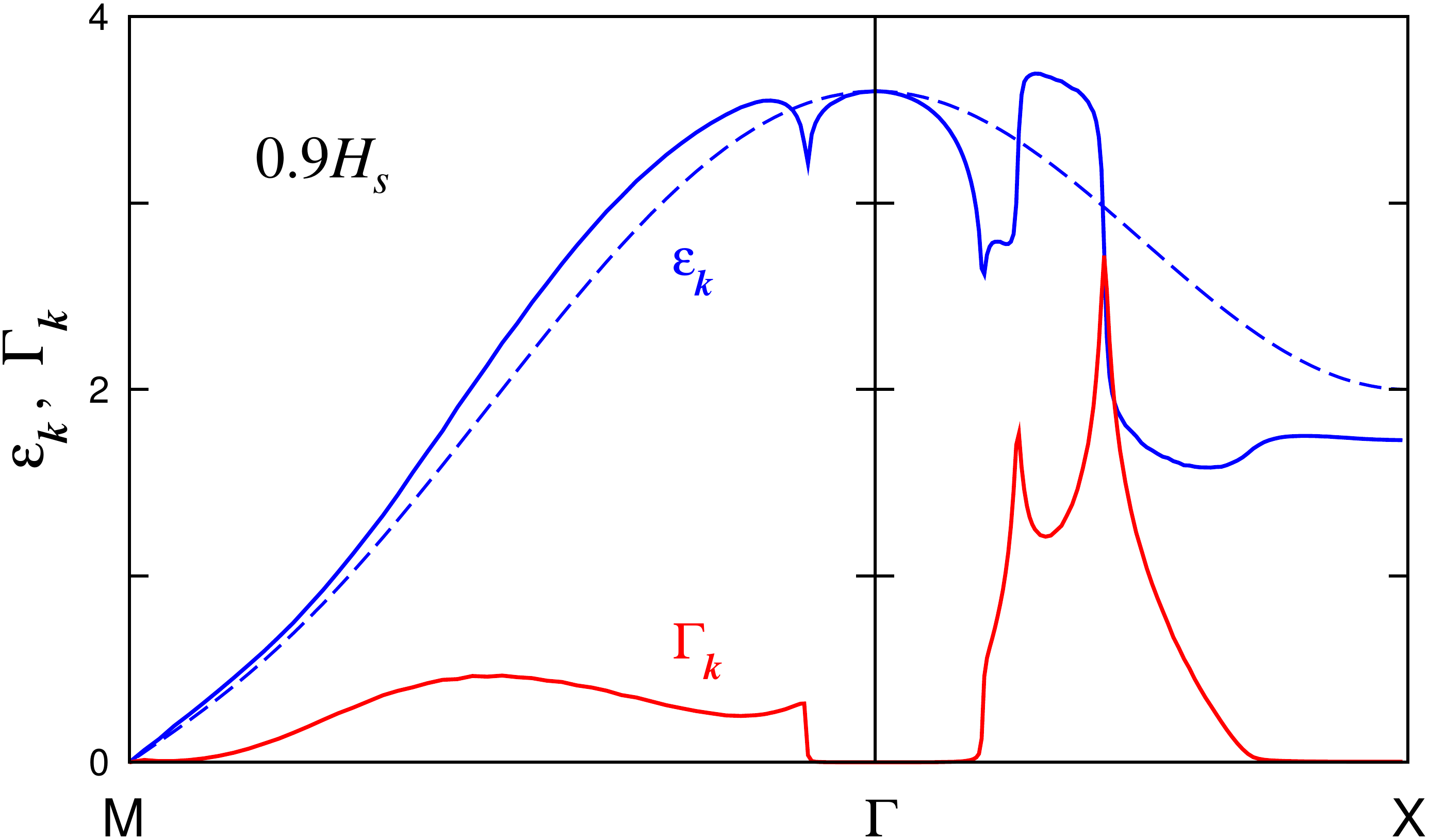}}
\caption{(color online).\ \
Magnon energy and decay rate in units of $J$  for the spin-$\frac{1}{2}$ square-lattice
antiferromagnet along the $M\Gamma X$ path at $H=0.9H_s$. The two upper curves
 show the harmonic $\varepsilon_{\bf k}$, dashed line, and the
renormalized magnon energy  $\bar\varepsilon_{\bf k}$, solid line.
The lower curve  gives the decay rate $\Gamma_{\bf k}$.
}
\label{fig:dispersion_high}
\end{figure}

Within the  spin-wave expansion, the $1/S$ renormalization for the magnon
spectrum is given by
\begin{equation}
\bar{\varepsilon}_{\bf k}=\tilde{\varepsilon}_{\bf k}+
{\rm Re}\,\Sigma_a({\bf k},\varepsilon_{\bf k})+\Sigma_b({\bf k},\varepsilon_{\bf k}),
\label{renorm_e}
\end{equation}
where
$\tilde{\varepsilon}_{\bf k} = \varepsilon_{\bf k} + \delta\varepsilon_{\bf k}$
already contains the Hartree-Fock corrections to the linear spin-wave dispersion
$\varepsilon_{\bf k}$ [see Eq.~(\ref{eq:cubic})], and the magnon damping is
$\Gamma_{\bf k}= -\textrm{Im}\,\Sigma_a({\bf k},\varepsilon_{\bf k})$ in the same
approximation. In Fig.~\ref{fig:dispersion_high} we present numerical results
for the renormalized magnon energy and damping for $H=0.9H_s$, $S=1/2$, and
for ${\bf k}$ along  the $M\Gamma X$ line, which crosses several threshold
boundaries according to Fig.~\ref{fig:decayregion_09}. The magnon energy exhibits
singularities that  are clearly related to spontaneous magnon decays: every spike
or jump in $\bar{\varepsilon}_{\bf k}$ is accompanied by a jump or a peak in the
decay rate $\Gamma_{\bf k}$. In agreement with Sec.~\ref{sec:kinematics_singular},
the logarithmic singularities occur on  a crossing of the singularity contour with
substantial density of two-magnon states, a minimum or a saddle point of the
continuum outlined in Fig.~\ref{fig:decayregion_09}.

Such singularities invalidate the usual $1/S$ expansion because the renormalized 
spectrum contains divergences. The encountered problem is generic and must be common
to a variety of the 2D models {\it regardless} of the value of the on-site spin $S$.
For systems with large spin the situation is especially aggravating as
the $1/S$ expansion changes from a reliable approach to one producing unphysical 
divergences, undermining any sensible comparison with experimental or numerical data.

Singular corrections to the spectrum obtained in the leading order require
a higher-order regularization, which depends on the form of $\varepsilon_{\bf k}$.
For example, two-roton decay in the superfluid $^4$He yields  logarithmic divergence 
of the one-loop diagram in Fig.~\ref{fig:diagrams}(a) near the decay threshold.
Using regularization  of the two-particle scattering amplitude, \textcite{Pitaevskii59} 
showed that such thresholds become termination points of the spectrum; this was later 
confirmed experimentally \cite{Glyde98}. However, this salient feature of the $^4$He 
spectrum is owing to the form of $\varepsilon_{\bf k}$, which ensures that rotons 
created in the decay process are well defined. The problem considered here is 
qualitatively different because magnons created in an elementary decay are also unstable. 
This can be seen from the evolution of the decay regions versus field in 
Fig.~\ref{fig:decayregion_safm}, which begins with the acoustic branch where the typical 
decay processes produce magnons  further within that region.

Such ``cascade''  decays imply the following.
First, the threshold and saddle-point singularities in
Eqs.~(\ref{logB}) and (\ref{logS})  are removed  by changing
$|\Delta k| \rightarrow \sqrt{(\Delta k)^2 + (\Gamma_{\bf q^*}/c_{\bf q^*})^2}$,
where $\Gamma_{\bf q^*}$ is the damping of the ``singular'' decay products and
$c_{\bf q^*}$ is related to their velocities \cite{Mourigal10}.
Second, the higher-order regularization treatment must include
damping of the  decay products.
Thus, as in the Sec.~IV.C, dressing of the bare propagators
in  Fig.~\ref{fig:diagrams}(a)  within the  SCBA should yield a regularized spectrum.
Having in mind the technical difficulty of this approach
we  used instead another scheme, referred to as imaginary-part SCBA (iSCBA).
This treats the magnon decay rate $\Gamma_{\bf k}$ self-consistently,
but neglects quantum corrections to $\varepsilon_{\bf k}$ altogether \cite{Mourigal10}.
Indeed,  away from singularities and for larger spins, $S\geq 1$,
renormalization of the spectrum $\delta\varepsilon_{\bf k}\sim {\cal O}(J)$
 is small compared to the bare energy $\varepsilon_{\bf k}\sim {\cal O}(SJ)$
 and can be neglected.
Although the intrinsic damping  is
of the same order $\Gamma_{\bf k}\sim {\cal O}(J)$,
it represents the most important qualitatively new feature of the spectrum acquired due to
spontaneous decays.

\begin{figure}[t]
\centerline{
\includegraphics[width=0.999\columnwidth]{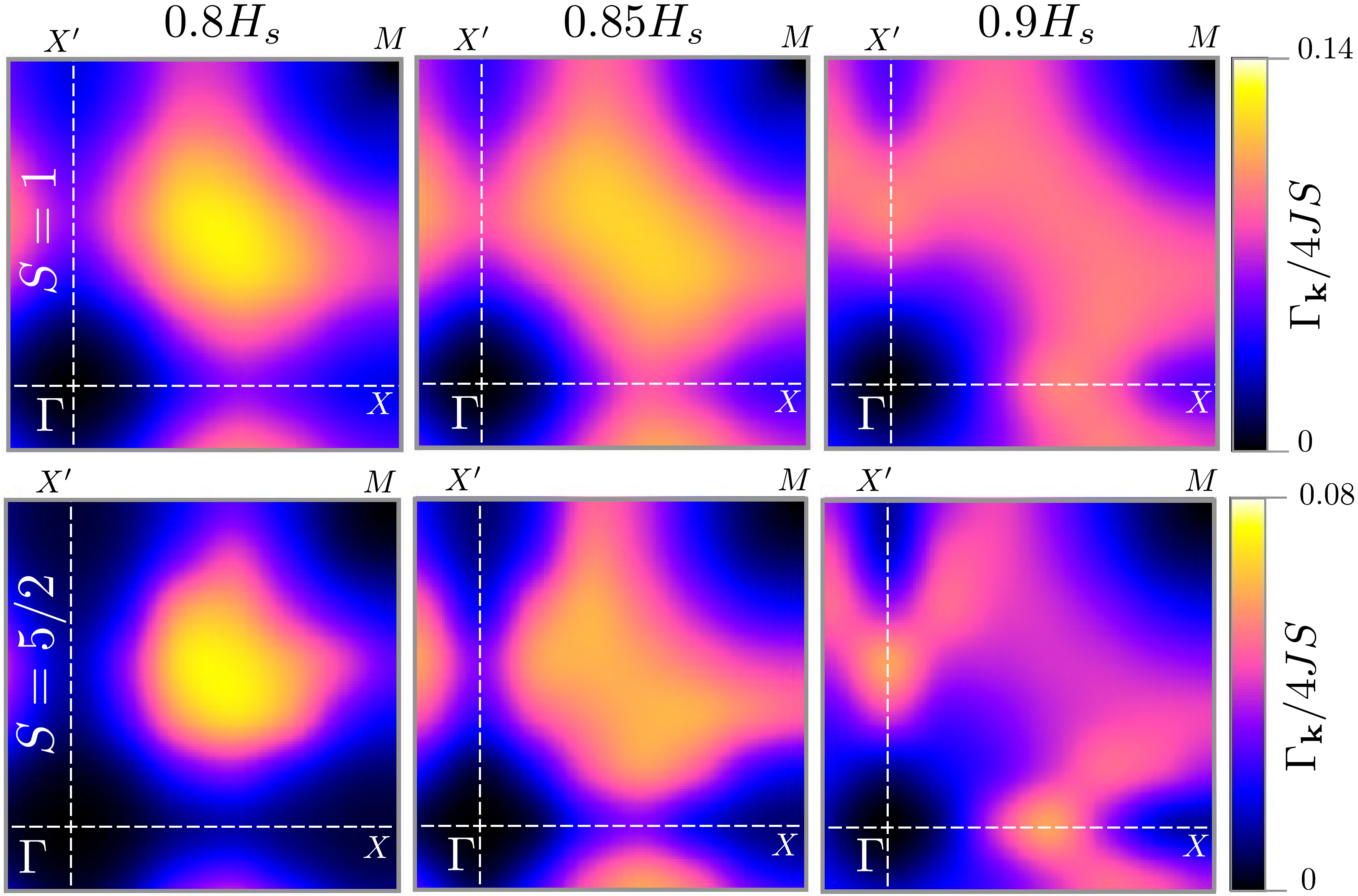}
}
\caption{(color online).\ \
Intensity maps of the magnon decay rate $\Gamma_{\mathbf{k}}$
calculated within the iSCBA  for several values of external magnetic field
and for two values of the spin:  $S=1$ (upper panels) and
$S=5/2$ (lower panels).
}
\label{fig:Gk_scba}
\end{figure}

In the simplest realization of the iSCBA we impose the following ansatz on
the Green's function:
\begin{equation}
G^{-1}({\bf k},\omega) =  \omega -\varepsilon_{\bf k} + i \Gamma_{\bf k} \ .
\label{Green1}
\end{equation}
This approximation  is equivalent to assuming
a Lorentzian shape of the quasiparticle peaks in the dynamical response.
Neglecting contributions from the source diagram,
the  self-consistent equation for the magnon decay rate $\Gamma_{\bf k}$ becomes
\begin{equation}
\Gamma_{\bf k} =
\frac{1}{2}\sum_{\bf q} \frac{\big|V_3^{(1)}({\bf q};{\bf k})\big|^2
(\Gamma_{\bf q}+\Gamma_{\bf k-q+Q})}
{(\varepsilon_{\bf k}-\varepsilon_{\bf q}-\varepsilon_{\bf k-q+Q})^2 +
(\Gamma_{\bf q}+\Gamma_{\bf k-q+Q})^2} \ .
\label{iSCBA}
\end{equation}
Such a scheme was previously applied to the
problem of phonon damping
in $^4$He \cite{Sluckin74} and in quasicrystals \cite{Kats05}.

One of the qualitative results of the
iSCBA approach can be immediately anticipated: since by dressing the diagram in
Fig.~\ref{fig:diagrams}(a) we implicitly take into account all $n$-particle
decay  processes,  the ${\bf k}$-region where magnons are damped is no longer confined to
the two-magnon decay region. In Fig.~\ref{fig:decayregion_safm} we provide
an indication of  that by showing the decay regions for the three-magnon processes (dashed lines).
These are generally weaker than the leading two-magnon decays, but their kinematics  is less restrictive
as for any given $H>H^*$ the two-magnon  decay area in Fig.~\ref{fig:decayregion_safm}
is enclosed by the three-magnon one.
In the limit of $n\to\infty$, the decay region extends farther and can cover
the entire  Brillouin zone in the high-field regime.

The same conclusion is reached if we consider broadening of the decay threshold
singularity by the finite lifetime of the final-state magnons. This broadening
implies that  the boundary in the momentum space between stable and damped magnons
is smeared out and the decay probability disappears gradually rather than in a
steplike fashion.

The numerical solution of the iSCBA Eq.~(\ref{iSCBA})
for two representative values of spin $S=1$ and $S=5/2$ and for several field values
is presented in Fig.~\ref{fig:Gk_scba}
in the form  of intensity maps of $\Gamma_{\bf k}/4JS$ vs ${\bf k}$  in the same part of the
Brillouin zone as in Fig.~\ref{fig:decayregion_09}.
As discussed, the singularities in $\Gamma_{\bf k}$
are washed out and finite damping develops
 outside of the original decay regions of Figs.~\ref{fig:decayregion_safm} and \ref{fig:decayregion_09}.
Comparison of the upper ($S=1$) and lower ($S=5/2$) rows   in Fig.~\ref{fig:Gk_scba}  reveals
similar patterns in the momentum distribution of $\Gamma_{\bf k}$.

\begin{figure*}[t]
\includegraphics[width=1.5\columnwidth]{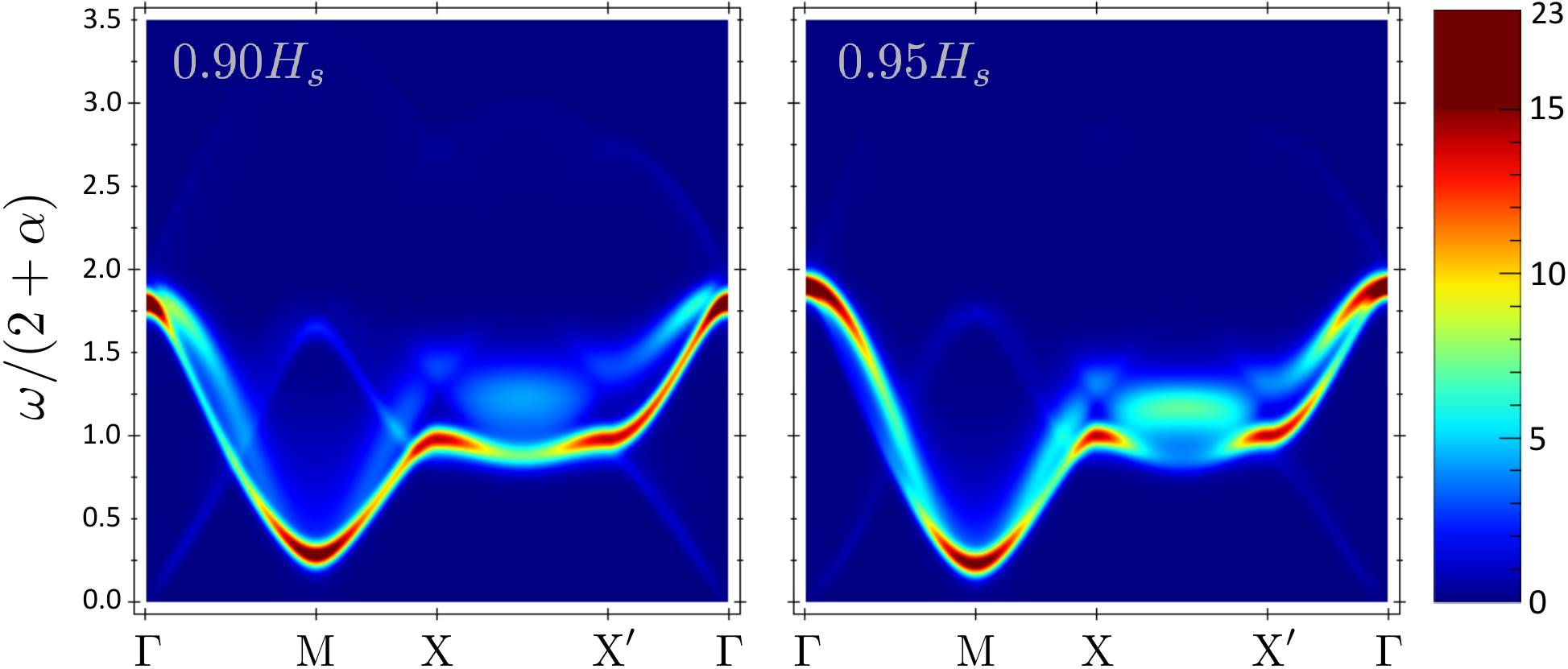}
\caption{(color online).
Intensity plots of the dynamical structure factor $S({\bf k},\omega)$,
for fields $0.9H_s$ and $0.95H_s$,  $S=1/2$, and
for the ${\bf k}$ path in the $k_z\!=\!0$ plane, along the
$\rm{\Gamma} \rightarrow \rm{M} \rightarrow \rm{X} \rightarrow \rm{X'} \rightarrow\Gamma$
line; see Fig.~\ref{regions} for notations.  From \textcite{Fuhrman12}.
}
\label{skw1}
\end{figure*}

Specifically, in lower fields, $H\alt 0.9H_s$, magnon damping
is most significant in the broad region around the $(\pi/2,\pi/2)$ point,
which is a consequence of the large phase-space volume
for the two-particle decays in this region.
In higher fields, the maxima in the decay rate shift to
the $\Gamma X$ $(\Gamma X')$ lines where the enhancement of damping is due to
 the remnants of  the strong logarithmic saddle-point singularities in Fig.~\ref{fig:dispersion_high}.
For $S\gg 1$ one can show that  $\Gamma_{\bf k}\sim {\cal O}(J)$
away from the former singular contours but there is a parametric enhancement of damping
$\Gamma_{\bf k}\sim {\cal O}(J\ln S)$  in their vicinity \cite{Chernyshev09a}.

To summarize, the self-consistently calculated magnon damping in the large-$S$
square-lattice antiferromagnet in strong field is free from singularities and
does not exceed $\Gamma_{\bf k}^{\rm max}\sim 0.7J\!-\!0.8J$.
In other words,  spin waves in large-$S$ systems acquire finite lifetimes at
$H>H^*$ but remain well-defined quasiparticles. The regularized singularities
lead to a parametric enhancement of magnon damping and produce distinct field
dependence and momentum dependence of $\Gamma_{\bf k}$.
The above results are in qualitative agreement with the inelastic
neutron-scattering data on the spin-$\frac{5}{2}$ layered square-lattice antiferromagnet
Ba$_2$MnGe$_2$O$_7$ \cite{Masuda10}, although more refined measurements are
necessary for a detailed comparison between theory and experiment.

\subsection{Decays in three dimensions}
\label{sec:3D}

In the two preceding sections we presented results of
self-consistent schemes that dealt with the unrealistic escape
of the quasiparticle peaks from the unrenormalized two-magnon
continuum and with singularities encountered in the $1/S$ expansion.
However, as noted in Sec.~\ref{sec:kinematics_singular}, the
decay-induced divergencies in $\Gamma_{\bf k}$ and $\bar{\varepsilon}_{\bf k}$
in 2D are replaced by a continuous albeit nonanalytic behavior in 3D.
Recent work by \textcite{Fuhrman12} demonstrated that
a moderate interlayer coupling mitigates 2D singularities and
represents an alternative to the above self-consistency schemes.
This approach allows one to obtain the excitation spectrum  using the standard spin-wave expansion,
{\it without} the use of self-consistency,
with no restrictions on the spectral shapes and at a fraction of  the computational cost.
As a specific example,  we  calculated the dynamical structure factor
of the  quasi-2D  $S\!=\!1/2$ square-lattice antiferromagnet
with moderate interlayer coupling  $J'/J\!=\!0.2$, which is relevant to one
of the prospective materials
for observing magnon decays, $\rm{(5CAP)_2 CuCl_4}$, where 5CAP is 2-amino-5-chloropyridinium
\cite{Coomer07}.

In Fig.~\ref{skw1}, we show these results in the form of intensity plots of the
dynamical structure factor, $S({\bf k},\omega)$ for $H=0.9H_s$ and $0.95H_s$
along a  representative path in the Brillouin zone; see Fig.~\ref{regions} for notations and
\textcite{Fuhrman12} for details.
In addition to the main features  that are directly related to the magnon spectral
function $A({\bf k},\omega)$, the out-of-plane transverse mode contribution is visible as
a shadow, shifted by the ordering vector ${\bf Q}=(\pi,\pi,\pi)$. We performed a
Gaussian convolution in the $\omega$ direction with a $\sigma$ width of $0.1 J$.
This step is intended to mimic the effect of a realistic experimental resolution, from which
we can also draw quantitative predictions of the relative strength of the quasiparticle and
the incoherent  part of the spectrum.

Complex spectral lineshapes, very much distinct from the conventional quasiparticle peaks,
are clearly visible in Fig.~\ref{skw1}. This demonstrates that the effects of spectral
weight redistribution and broadening due to spontaneous decays are substantial and should
be readily observed in experiment. The structure factor
exhibits a variety of unusual features, including the  double-peak lineshape
for the $\Gamma M$ direction discussed previously.
Along some of the ${\bf k}$ directions for $H=0.9H_s$,
the renormalized quasiparticle peaks escape the unrenormalized two-magnon continuum and survive
despite being formally within the decay region shown in  Fig.~\ref{regions}.
Yet these peaks are accompanied
by continuumlike subbands that accumulate significant weight.
Upon increase of the field,
the continuum overtakes the single-particle branches in large regions of the Brillouin zone,
washing them out and creating more double-peak structures, e.g.,  very distinctly
along the $X'\Gamma$ direction. There is
a particularly spectacular advance of the continuum along the $XX'$ line, where
magnon broadening  is most intense with dramatic spectral weight transfer to higher energies.

The overall landscape of the high-field dynamical structure factor shown here
is diverse and intriguing. Our results, consistent with prior numerical and
analytical studies, provide a comprehensive illustration of its details.
Inelastic neutron-scattering measurements on suitable quantum antiferromagnets
will be important to elucidate the accuracy of our theoretical results.

\section{Zero-field decays}
\label{sec:tafm}

In this section we discuss spontaneous decays in noncollinear antiferromagnets
in zero magnetic field. As explained in Sec.~\ref{sec:anharmonic}, the
noncollinear magnetic structures {\it necessarily} lead to the coupling of
transverse and longitudinal spin fluctuations that translates into cubic
anharmonicities in the effective magnon Hamiltonian. This opens the door to
spontaneous two-magnon decays without the assistance of an external field, provided
the kinematic conditions of Sec.~\ref{sec:kinematics} are fulfilled. Here we
concentrate on the case of  {\it spiral} Heisenberg antiferromagnets, in which
the noncollinear ground state  spontaneously breaks the SO(3) symmetry of
the spin Hamiltonian. We show that the kinematic conditions that allow two-magnon
decays are {\it always} fulfilled in such a case, owing to the existence of
three acoustic branches in the excitation spectra. Next we elucidate these
features and consider magnon decays in the spin-$\frac{1}{2}$ triangular-lattice
antiferromagnet in more detail.

Our subsequent discussion does not include noncollinear magnetic
structures induced by the anisotropic Dzyaloshinskii-Moriya interactions,
which have attracted a lot of interest because of experimental observation of
the skyrmion lattices \cite{Rossler06}. We also leave aside the other cases
of noncollinear antiferromagnets, such as kagom\'{e}-lattice potassium jarosite
\cite{Matan06} and anisotropic triangular-lattice antiferromagnets,
discussed by \textcite{Chernyshev09a}, although many qualitative aspects of our
consideration below remain valid for them. Yet another class of zero-field
spontaneous decays that do not require noncollinearity, the three-magnon decays
in anisotropic ferromagnets \cite{Villain74,Stephanovich11}, is also left out.

\subsection{General discussion}
\label{sec:helix}

Noncollinear spin structures in isotropic antiferromagnets appear as a
consequence of magnetic frustration: either because of the geometry
of the nearest-neighbor bonds as, {\it e.g.}, for a triangular lattice, or due
to competing exchange interactions beyond the nearest-neighbor shell.
Here we consider a general case of the Heisenberg
Hamiltonian
\begin{equation}
\hat{\cal H} = \sum_{\langle ij\rangle} J({\bf r})\; {\bf S}_i\cdot {\bf S}_j
\label{Hspiral}
\end{equation}
where spins occupy the sites of a Bravais
lattice and ${\bf r}\equiv {\bf r}_j-{\bf r}_i$. The lowest-energy classical state of this model
is a spin helix with a wave vector $\bf Q$, which is found by minimizing the Fourier transform of
the exchange energy $J_{\bf q} = \sum_{\bf r} J({\bf r})\cos({\bf q\cdot r})$.

The spin-wave theory has been applied to spiral magnetic structures in a number of  works;
see, for example, \textcite{Chubukov84,Ohyama93}.
Similarly to the procedure discussed in Sec.~\ref{sec:safm_SWT}, the classical spin configuration
defines the rotating local basis for spin quantization, which is  related to the laboratory
frame  via
\begin{eqnarray}
&& S_i^{x_0} = S_i^z \sin{\bf Q}\cdot{\bf r}_i + S_i^x\cos{\bf Q}\cdot{\bf r}_i \ , \quad
S_i^{y_0} = S_i^y \ , \nonumber \\
&& S_i^{z_0} = S_i^z \cos{\bf Q}\cdot{\bf r}_i - S_i^x\sin{\bf Q}\cdot{\bf r}_i  \ .
\label{Szx}
\end{eqnarray}
The subsequent $1/S$ expansion follows standard steps (see Sec.~\ref{sec:safm_SWT})
and yields the spin-wave dispersion in the harmonic approximation:
\begin{equation}
\varepsilon_{\bf k} = S\sqrt{\left(J_{\bf k} - J_{\bf Q}\right)
\left({\textstyle \frac{1}{2}}J_{\bf Q+k}+{\textstyle \frac{1}{2}}J_{\bf Q-k}-J_{\bf Q}\right)} \ .
\label{tri_ek}
\end{equation}
One can see that for an arbitrary form of $J_{\bf q}$ the excitation energy vanishes at three
wave vectors: ${\bf k = 0}$ and $\pm{\bf Q}$. The existence  of these three
Goldstone modes  follows from the complete breaking
of the rotational SO(3) symmetry in the noncollinear ground state \cite{Andreev80,Dombre89}.
For a planar magnetic structure, spin waves with ${\bf k} = \pm{\bf Q}$
correspond to the out-of-plane oscillation modes related by
symmetry, whereas the ${\bf k = 0}$ mode represents the in-plane
oscillations of the spin spiral.
Overall there are three acoustic branches,
shown schematically in Fig.~\ref{fig:acoustic}(c), with two distinct
spin-wave velocities $c_0$ and $c_Q$.

As was noted, coupling between one- and
two-magnon states is inevitable in noncollinear structures.
Hence, to verify the presence or absence of spontaneous decays in spiral
antiferromagnets one only needs to analyze the kinematic constraints following
from the energy conservation (\ref{energy_conserv}).
Our discussion in Sec.~\ref{sec:kinematics_acoustic} suggests that
in a system with multiple acoustic modes
the fast low-energy magnons can always decay into slow ones. Using Eq.~(\ref{tri_ek})
one can obtain the velocities of the two acoustic modes as
\begin{eqnarray}
&& c_0 = S\sqrt{A(J_0 -J_ {\bf Q})}\ , \quad
A = \frac{1}{2}\left.\nabla^2_{\bf k}J^{_{}}_{\bf k}\right|_{\bf k=Q} \ ,
\nonumber \\
&& c_Q =  S\sqrt{A\left({\textstyle \frac{1}{2}}J_0 +{\textstyle \frac{1}{2}}J_{2\bf Q}-J_{\bf Q}\right)}\ .
\label{c0_cQ}
\end{eqnarray}
Since the ordering wave vector of a spiral structure has to satisfy the inequality
$2{\bf Q}\neq {\bf G}_i$, the ${\bf G}_i$'s being  reciprocal lattice vectors,
the two velocities in Eq.~(\ref{c0_cQ}) are generally different.
If $J_0 >J_{2\bf Q}$, the  in-plane mode is faster: $c_0>c_Q$.
For example, this relation holds if all exchange constants are antiferromagnetic,
$J({\bf r})>0$.  In such a case  the decay channel
${\bf k} \to ({\bf Q}+{\bf q}) + (-{\bf Q}+{\bf q'})$ is always open for sufficiently small ${\bf k}$.
Since for the linearized spectra decays would be permitted for any ${\bf k}$, it is the
short-range behavior of $J_{\bf q}$ that limits the size of the decay region, which
typically occupies an extended region of the Brillouin zone including  the $\Gamma$ point.
Generally, acoustic branches emanating
from the $\pm{\bf Q}$ points remain undamped for an incommensurate ${\bf Q}$,
although, see the discussion of the triangular-lattice case in Sec.~\ref{sec:triangular}.
If, on the other hand,  $J_0<J_{2\bf Q}$, which happens when some bonds
are ferromagnetic, out-of-plane modes are faster: $c_0<c_Q$.  Now the decay
channel is switched to $({\bf Q}+{\bf k}) \to ({\bf Q}+{\bf q}) + {\bf q'}$ for
small ${\bf k}$ such that the decay region surrounds  the $\pm{\bf Q}$ points,
while magnons in the center of the Brillouin zone remain undamped.

In this discussion we disregarded  possible anisotropy of
the spin-wave velocities.  However, since
 both $c_0$ and $c_Q$  have the same angular dependence via $A$ in Eq.~(\ref{c0_cQ}),
the anisotropy plays only a secondary role, which does not change the
qualitative conclusion about the magnon decay  region.

A finite lifetime of magnetic excitations was noted
in several studies of noncollinear antiferromagnets within the higher-order spin-wave theory
\cite{Ohyama93,Veillette05,Dalidovich06,Starykh06}.
The generality of the phenomenon of spontaneous magnon decays in spiral
antiferromagnets was emphasized in our  work \cite{Chernyshev06},
in which we studied in detail the triangular-lattice Heisenberg
antiferromagnet considered next.

\subsection{Triangular-lattice antiferromagnet}
\label{sec:triangular}

\begin{figure}[t]
\centerline{
\includegraphics[width=0.7\columnwidth]{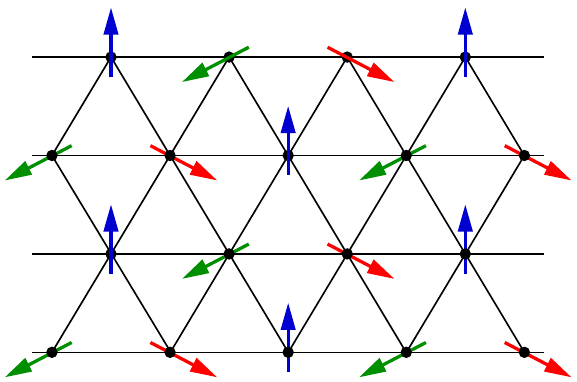}}
\caption{(color online).\ \
The ordered 120$^\circ$ spin structure of the
triangular-lattice Heisenberg antiferromagnet.
}
\label{fig:120_structure}
\end{figure}

The semiclassical $S\gg 1$ triangular-lattice antiferromagnet orders in the
120$^\circ$ spin structure shown in Fig.~\ref{fig:120_structure}. For the case
of $S=1/2$, it was proposed that the enhanced quantum fluctuations may completely
destroy the long-range antiferromagnetic order \cite{Anderson73,Fazekas74}.
However, the spin-wave calculations suggested that the 120$^\circ$ state is
stable for $S=1/2$, albeit with a significantly reduced ordered moment
\cite{Jolicoeur89,Miyake92}. The exact diagonalization of small clusters
\cite{Bernu94} together with the more recent QMC \cite{Sorella99}, series-expansion
\cite{Zheng06c}, and density-matrix renormalization group studies \cite{White07}
confirmed the stability of the 120$^\circ$ spin-structure for the spin-$\frac{1}{2}$ case
and yielded $\langle S\rangle \approx 0.20$ for the ordered moment.

The series-expansion results for the excitation spectrum of the spin-$\frac{1}{2}$
triangular-lattice antiferromagnet \cite{Zheng06a} deviate substantially from
the harmonic spin-wave theory with the overall band-narrowing by $\sim\! 50$\%
and additional rotonlike minima. These effects were shown to result from strong
magnon-magnon interactions by \textcite{Starykh06} and \textcite{Zheng06c}, who
demonstrated that the first-order $1/S$ renormalization gives a spectrum in
qualitative agreement with the series-expansion data. The importance of decays
in the magnon spectrum of the triangular-lattice antiferromagnet was pointed
out by \textcite{Chernyshev06}.

\begin{figure}[b]
\centerline{\includegraphics[width=0.65\columnwidth]{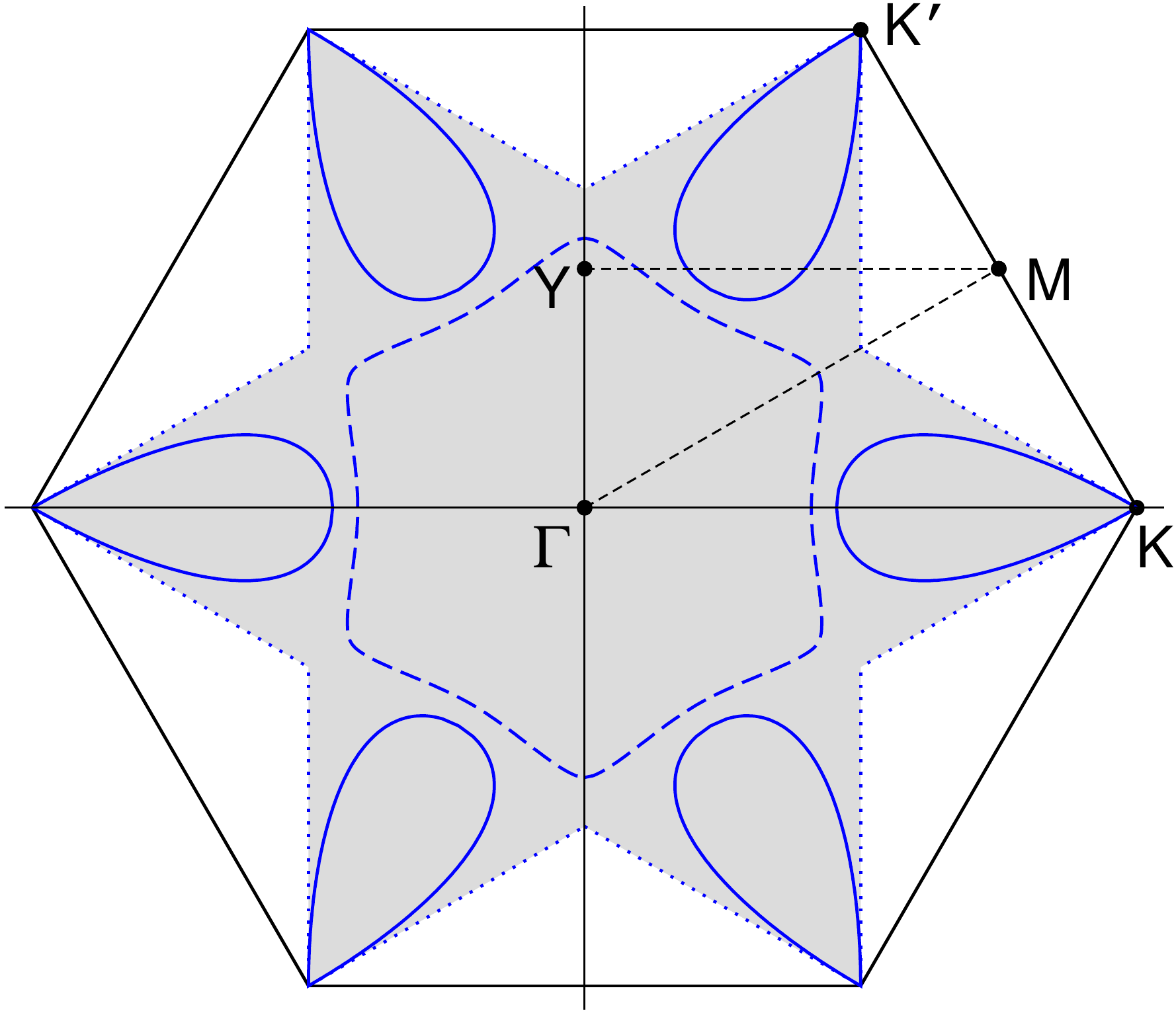}}
\caption{(color online).\ \
Brillouin zone of the triangular lattice. The magnon decay
region is shown as a shaded area.  Solid and dashed lines indicate saddle-point
van Hove singularities of the two-magnon continuum.}
\label{fig:decayregion_tafm}
\end{figure}

We now proceed with a brief description of the spin-wave results for
the dynamics of the triangular-lattice antiferromagnet, referring intrested readers
to  \textcite{Chernyshev09a} for a detailed review.
We consider the Heisenberg Hamiltonian (\ref{Hspiral}) with the exchange interaction
between the nearest-neighbor sites $J({\bf r})\equiv J\delta_{|{\bf r}|,1}$.
In this case, the Fourier transform of the exchange matrix is
$J_{\bf q}=6J\gamma_{\bf q}$ with
$\gamma_{\bf q} = \frac{1}{3}[\cos q_x + 2 \cos(q_x/2)\cos(\sqrt{3}q_y/2)]$.
The ordering wave vector of the 120$^\circ$ spin structure
${\bf Q}=(4\pi/3,0)$  is commensurate with the reciprocal lattice and is located at the
$K$ point in the corner of the Brillouin zone; see Fig.~\ref{fig:decayregion_tafm}.

The magnon dispersion in the harmonic approximation is
\begin{equation}
\varepsilon_{\bf k} =
3JS \sqrt{(1-\gamma_{\bf k})(1+2\gamma_{\bf k})} \ .
\label{Ek_tafm}
\end{equation}
In full agreement with the  general picture outlined above,
the excitation spectrum has three Goldstone modes at ${\bf k=0},\pm{\bf Q}$
with distinct spin-wave velocities $c_0$ and $c_Q$. In the harmonic
approximation their ratio is $c_0/c_Q=\sqrt{2}$, with the leading $1/S$ renormalization
reducing it to $c_0/c_Q\approx 1.156$ in the $S=1/2$ case \cite{Chubukov94}.
Another notable feature of the dispersion in Eq.~(\ref{Ek_tafm}) is
the varying convexity of the acoustic branch in the vicinity of the ${\bf Q}$ points
\begin{equation}
\varepsilon_{{\bf Q}+{\bf k}}\approx c_Q k(1-\alpha_\varphi k)\ , \qquad
\alpha_\varphi \sim \cos 3\varphi\ ,
\label{wQ}
\end{equation}
which  changes its sign depending on the direction of ${\bf k}$.

Using the harmonic expression for $\varepsilon_{\bf k}$ we obtain the magnon
decay region of the triangular-lattice antiferromagnet as given by the shaded
hexagram in Fig.~\ref{fig:decayregion_tafm}. The decay threshold boundary is
determined solely by the emission of an acoustic magnon with ${\bf q}\to \pm{\bf Q}$;
see Eq.~(\ref{acousticQ}), shown in Fig.~\ref{fig:decayregion_tafm} by dotted
lines. As anticipated above, the decay region includes an extended neighborhood
of the $\Gamma$ point. Interestingly, the decay
region also extends all the way to the $K$, $K'$ points ($\equiv\pm{\bf Q}$). This  is
a consequence of varying convexity of the corresponding acoustic branch (\ref{wQ})
together with the commensurability of the ordering wave vector
$3{\bf Q}={\bf G}$. As a result, the kinematic
conditions are satisfied for a decay from the steeper side
of the energy cone with $\bf k\to Q$ into two magnons with $\bf q,q'\to -Q$.
Thus, magnons near $K$ or $K'$ points are unstable only in a certain range
of angles, in agreement with Fig.~\ref{fig:decayregion_tafm}.

In clear contrast with the  field-induced decays (see Fig.~\ref{fig:decayregion_safm}),
the decay region of the triangular-lattice antiferromagnet is not extended by the
three-magnon decays. This implies that the cascade multimagnon decays do not have
a tendency to spread over the  Brillouin zone and that the decay threshold boundary
remains sharp. The boundary can change due to renormalization of the harmonic
dispersion (\ref{Ek_tafm}), but the corresponding shifts appear to be not very
significant \cite{Chernyshev09a}.

Last, the solid and dashed lines in  Fig.~\ref{fig:decayregion_tafm} show
the threshold boundaries for the emission of two equivalent and inequivalent
magnons, respectively; see Sec.~\ref{sec:kinematics_threshold}. Both sets of
contours lie completely within the decay region and thus  correspond to the
saddle points of the two-magnon continuum, which result in the  logarithmic
peaks in the magnon decay rate $\Gamma_{\bf k}$; see
Sec.~\ref{sec:kinematics_singular}.

The  $1/S$ renormalization of the magnon spectrum  of the triangular-lattice
antiferromagnet is calculated in the same way as in the case of the square-lattice
antiferromagnet, discussed in Sec.~\ref{sec:field_decays}. The minor technical
simplification is in the absence of the angle renormalization in the present case
due to the high symmetry of  the 120$^\circ$ state. Aside from this detail, the
renormalized ``on-shell'' spectrum is given by Eq.~(\ref{renorm_e}) with the
self-energies $\Sigma_a({\bf k},\varepsilon_{\bf k})$ and
$\Sigma_b({\bf k},\varepsilon_{\bf k})$ [see Figs.~\ref{fig:diagrams}(a,b)],
and the Hartree-Fock correction from the quartic terms (\ref{eq:cubic}). All
relevant expressions can be found in \textcite{Chernyshev09a};
see also \textcite{Starykh06}  for a somewhat different procedure of deriving
the $1/S$ renormalization.

\begin{figure}[t]
\begin{center}
\includegraphics[width = 0.95\columnwidth]{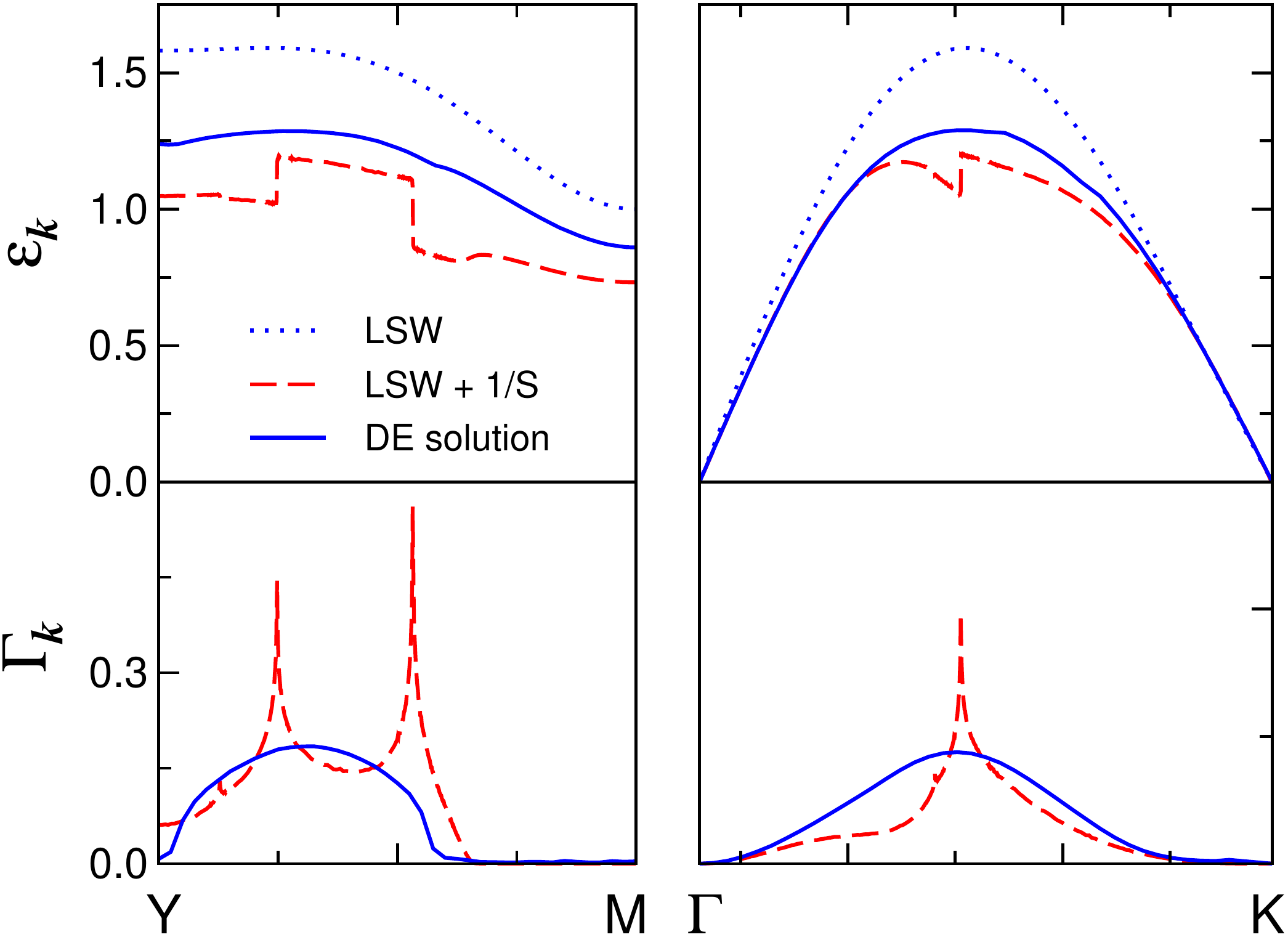}
\end{center}
\caption{(color online).\ \
Magnon energy and decay rate in units of $J$ for the spin-$\frac{1}{2}$ triangular-lattice
antiferromagnet. Dotted and  dashed lines are the harmonic  and the $1/S$ renormalized results,
respectively. Solid lines are solutions of the Dyson equation (\ref{DE_SC1})
in the complex plane.
}
\label{fig:Ek_tafm}
\end{figure}

The magnon energy and the decay rate obtained in the $1/S$ approximation (\ref{renorm_e})
are shown by dashed lines in Fig.~\ref{fig:Ek_tafm} along the representative ${\bf k}$
directions. In addition to the singularities, similar to the square-lattice case
in Fig.~\ref{fig:dispersion_high}, there is a significant reduction of the magnon bandwidth.
This substantial downward energy renormalization is due to strong repulsion of one- and
two-magnon states that are coupled by the cubic anharmonicities.
One can expect such interactions to produce similarly strong downward shifts
of the excitation energies in the other spiral antiferromagnets.
For example, in the anisotropic  $XY$ limit of the $S=1/2$ triangular-lattice antiferromagnet,
reduction of the bandwidth due to magnon interactions is even stronger  and exceeds 50\%
 of the harmonic value \cite{Chernyshev09a}, a magnitude unheard of in the collinear
 antiferromagnets.

Finally, the magnon damping, the hallmark of magnon decays in spiral antiferromagnets,
 is substantial throughout much of the decay region of Fig.~\ref{fig:decayregion_tafm}.
As noted, the channel that determines the decay threshold boundary in the present case
is the emission of the acoustic magnon. Because of that,  the threshold behavior of the
decay rate near the boundary in Fig.~\ref{fig:Ek_tafm} is not singular,
$\Gamma_{\bf k} \sim (k-k_b)^2$ \cite{Chernyshev06}, in contrast with the square-lattice
case (see Fig.~\ref{fig:dispersion_high}), where the threshold singularities are due to
minima in the two-magnon density of states.

There is another important difference in the dynamics of two-particle decays
in these two cases. Magnons emitted in an elementary decay process
in the square-lattice antiferromagnet are also unstable.
Allowing for a finite lifetime of the decay products using self-consistent
dressing of Fig.~\ref{fig:diagrams}(a) gave us divergence-free magnon damping in
Sec.~\ref{sec:field_decays}. For the triangular-lattice antiferromagnet this
approach may not work as the emitted magnons are often stable. In particular, this
is true for the vicinity of the logarithmic singularities indicated by solid lines in
Fig.~\ref{fig:decayregion_tafm}: the saddle-point momenta of the two-magnon continuum,
which are responsible for the singular behavior, can
be shown to fall outside the decay region. Hence, at the saddle points,
the logarithmic divergence of the one-loop diagrams will persist
even for the renormalized spectrum and the singularities seem to remain
essential.  Thus, the construction of a renormalized
theory seems to require a summation of an infinite series of
diagrams that contain leading-order divergences, as in
the Pitaevskii treatment of the spectrum termination problem \cite{Pitaevskii59}.
Such a procedure is technically problematic as it goes beyond the standard $1/S$
spin-wave expansion and  allows only for qualitative results \cite{Chernyshev09a}.

This seemingly hopeless situation is resolved if we note that the logarithmic
singularity occurs for a magnon  inside the decay region, which means that
there is always a ``background'' of the nonsingular decays in $\Gamma_{\bf k}$.
In that case, one has to allow for a finite lifetime of the {\it initial}
magnon while magnons created during the decay process may remain stable.
Technically, the corresponding procedure amounts to solving the ``off-shell''
Dyson equation for complex $\varepsilon=\bar\varepsilon_{\bf k}-i\Gamma_{\bf k}$,
preserving causality by substituting $\varepsilon\to\varepsilon^*$ in the argument
of the self-energy \cite{Chernyshev09a}. Then the Dyson equation is reduced to the system
\begin{eqnarray}
\bar\varepsilon_{\bf k} = \varepsilon_{\bf k} +
\textrm{Re}[\Sigma({\bf k},\varepsilon^*)]\, , \ \ \
\Gamma_{\bf k} =
-\textrm{Im}[\Sigma({\bf k},\varepsilon^*)]\ .
\label{DE_SC1}
\end{eqnarray}
The off-shell approach avoids complications related to higher-order multiloop
diagrams and naturally regularizes the decay singularities in $\Gamma_{\bf k}$
and $\bar{\varepsilon}_{\bf k}$. Results obtained by numerically solving Eq.~(\ref{DE_SC1})
are shown in Fig.~\ref{fig:Ek_tafm} by solid lines.
The decay rates remain significant  even in the absence of logarithmic peaks
reaching $\Gamma_{\bf k}/\bar{\varepsilon}_{\bf k} \sim 0.15$.

\section{Damping of low-energy magnons}
\label{sec:damping_lowE}

The long-wavelength limit is where one usually hopes to find universal
behavior of the dynamical properties. It is also the regime where decay rates
can be calculated perturbatively because of the smallness of interaction
among low-energy excitations and due to reduction of the phase-space volume
available for decay processes. In this section we discuss a few general
scenarios for decays of the long-wavelength excitations with the goal of
identifying universal dependencies of $\Gamma_{\bf k}$ on ${\bf k}$ and
complement it with particular examples from the triangular- and
square-lattice antiferomagnets considered above.

As discussed in Sec.~\ref{sec:kinematics_acoustic}, there are several
typical situations for the overlap of the single-particle acoustic branch with
the two-particle continuum that allows for energy conservation to be
fulfilled in the decay process. Here we do not repeat this classification
and assume that such conditions are always satisfied.
Generally, the decay rate is given by
\begin{equation}
\Gamma_{\bf k} \sim \sum_{\bf q} |V_{{\bf k},{\bf q}}|^2
\delta(\varepsilon_{\bf k}-\varepsilon_{\bf q}-\varepsilon_{\bf k-q}) \ ,
\label{Gamma_app}
\end{equation}
where $V_{{\bf k},{\bf q}}$ is the decay vertex.
Because of energy conservation, the upper limit for the momentum
of created quasiparticles should be of the order of $k$.
Then, for the linear spectrum $\varepsilon_{\bf k}\sim k$, a naive
power counting suggests
\begin{equation}
\label{naive0}
\Gamma_{\bf k} \sim k^{D-1} |V_k|^2 \ ,
\end{equation}
where $k^D$ comes from the $D$-dimensional phase space, $1/k$ is due
to the reduction of that space to the decay surface from energy conservation,
and $V_k$ is the typical amplitude of the cubic vertex on the decay surface.
We assume that the cubic vertex  at small momenta follows the standard
``hydrodynamic'' form \cite{StatPhysII}: $V_{\bf k,q}\propto \sqrt{kqq'}$,
where $q'=|{\bf k}-{\bf q}|$, and that in a typical decay process all the momenta
are of the same order  $q, q'\sim k$. This makes $|V_k|^2\propto k^3$ and yields
a seemingly universal power law for the decay rate of an arbitrary acoustic mode:
\begin{equation}
\label{naive}
\Gamma_{\bf k} \sim k^{D+2} \ .
\end{equation}
For $D=3$, $\Gamma_{\bf k}\sim k^5$, which matches the well-known result for
the decay of a phonon with a convex dispersion curve \cite{Beliaev58}, but, as
we see shortly, only coincidentally.

In reality, the situation is more delicate and possible power-law exponents for
$\Gamma_{\bf k}$ vary depending on physical details. In the case of a single
weakly nonlinear acoustic branch, relevant to phonons in $^4$He,
$\varepsilon_{\bf k}=c k +\alpha k^3$, decays are allowed only for the positive
curvature $\alpha > 0$; see Sec.~\ref{sec:kinematics_acoustic}.
A long-wavelength phonon decays into a pair of quasiparticles with
their momenta directed in a narrow solid angle along the direction of the initial
momentum. The apex angle of the decay cone scales as $\theta\sim k$ such that
the phase space is $k^{2D-1}$ instead of the naive $k^D$. Then the restriction
from the energy conservation gives $1/k\theta^2 \sim 1/k^3$ instead of $1/k$
as considered previously. Altogether, for the case of the cubic upward curvature
of the spectrum, the answer is universal [see also \textcite{Kreisel08}]:
\begin{equation}
\Gamma_{\bf k} \sim k^{2D-1},
\label{less_naive}
\end{equation}
which indeed yields $\Gamma_{\bf k} \sim k^5$ in 3D \cite{StatPhysII,Beliaev58},
but  the answer for 2D is  $\Gamma_{\bf k} \sim k^3$, different from Eq.~(\ref{naive}).
The 2D result applies to the square-lattice antiferromagnet in the strong field
considered in Sec.~\ref{sec:field_decays}.

If several acoustic modes with different velocities are present, the fast
quasiparticle can always decay into two slow excitations
(see Sec.~\ref{sec:kinematics_acoustic}).  This situation is simpler than in
the previous case since the nonlinearity of the spectra plays no role.
The phase-space factor becomes $k^{D-1}$ now, in agreement with the
naive consideration above. Physical realizations of this scenario
include decays of the longitudinal phonon into two transverse ones in solids
\cite{Kosevich} as well as the decay of the ${\bf k\rightarrow 0}$ into two
${\bf k}\rightarrow\pm {\bf Q}$ magnons in the triangular-lattice Heisenberg
antiferromagnet considered in Sec.~\ref{sec:tafm},
a case also pertinent to other spiral antiferromagnets.
Interaction between phonons in crystals obeys the conventional
asymptotic behavior $V_{\bf k,q}\propto \sqrt{kqq'}$ and, consequently,
$|V_k|^2\propto k^3$. Therefore, our initial naive power-counting result in Eq.~(\ref{naive})
is, actually, valid for them.
However, in the case of the triangular-lattice antiferromagnet, the result is yet
different from that in Eq.~(\ref{naive}) because the three-magnon
vertex  is anomalous and scales as
$V_{\bf k,Q+q}\propto (q'-q)\sqrt{k/qq'}$ for small $k$ \cite{Chernyshev09a}.
For a typical decay process the initial and final momenta are of the same order,
$q,q' \sim k$, giving
$|V_{\bf k,Q+q}|^2 \sim k$ instead of $k^3$.
Altogether, for noncollinear Heisenberg antiferromagnets
in $D$ dimensions this yields
\begin{equation}
\Gamma_{\bf k} \sim k^{D} \ .
\label{Gamma0}
\end{equation}
Direct analytic calculation for long-wavelength magnons in the triangular
antiferromagnet gives a decay rate with the correct exponent
$\Gamma_{\bf k} \approx  0.79 k^2$.

The decay vertex for the ${\bf k}\!\rightarrow\!{\bf Q}$ magnon has
the conventional scaling $V_{\bf Q+k,-Q+q}\!\propto\!
\sqrt{kqq'}$, so the decay
probability is $|V_k|^2\!\sim\!k^3$. However, due to
a constraint on the angle between ${\bf k}$ and ${\bf q}$,
the decay surface in ${\bf q}$ space is a cigar-shaped ellipse
with length $\sim\!k$ and width $\sim\!k^{3/2}$. That
causes the restricted phase volume of decays to scale as $k^{(3D-5)/2}$ and
leads to a nontrivial ${\bf k}$ dependence of the decay rate:
\begin{equation}
\Gamma_{\bf k} \sim k^{(3D+1)/2} \ .
\label{GammaQ}
\end{equation}
Numerically, in the triangular antiferromagnet $\Gamma_{\bf k}\approx 1.2Jk^{7/2}$
along the $K\Gamma$ line, in agreement with Eq.~(\ref{GammaQ}).
Away from this direction,  damping exhibits
an anomalous angular dependence $\Gamma_{\bf k} \sim k^{7/2}/(\cos 3\varphi)^{3/2}$. It diverges upon
approaching the decay threshold boundaries at $\varphi=\pm\pi/6$ in Fig.~\ref{fig:decayregion_tafm}
and thus requires a self-consistent regularization.

Another case of the singular behavior of the decay rate in the long-wavelength
limit occurs in the consideration of the square-lattice antiferromagnet in high fields, Sec.~\ref{sec:field_decays}.
Here we outline an approach to regularizing such divergences \cite{Mourigal10}.
Using the asymptotic form of the magnon dispersion in (\ref{acoustic}), the decay vertex
$V_3^{(1)}({\bf q}; {\bf k})\propto \sqrt{kqq'}$, and the Born expression for the decay
rate in (\ref{Gamma2_k}) we obtain
\begin{equation}
\Gamma_{\bf k} = \frac{3J}{16\pi} \, \tan^2\!\theta\;
\sqrt{\frac{c}{6\alpha}}\; k^3 \ ,
\label{Gkasympt}
\end{equation}
in agreement with Eq.~(\ref{less_naive}) with expressions for $c$ and $\alpha$
provided in Eq.~(\ref{E_acoust}).
An important observation is that  the nonlinearity of the
spectrum $\alpha$ vanishes at the decay threshold field $H\rightarrow H^*$,
and, as a result, the coefficient in front of $k^3$  in Eq.~(\ref{Gkasympt}) diverges.
Such nonanalytic behavior is nothing but the long-wavelength version of the
decay threshold singularities discussed in Sec.~\ref{sec:kinematics_singular}.
The self-consistent regularization (\ref{iSCBA}) should  remain applicable in this limit.
In order to obtain the renormalized ${\bf k}$ dependence of the damping,
we substitute the general power-law ansatz for
the decay rate $\Gamma_{\bf k} = \beta k^n$, and assume that
the damping is much larger than the vanishing nonlinearity
but still much smaller than the magnon energy
$\alpha k^3 \ll \beta k^n \ll ck$.  In this case, the apex angle of the decay cone
scales as $\varphi^2 \sim k^{n-1}$ with $q\sim k$, which makes the decay vertex
$V_3^{(1)}({\bf q};{\bf k})\propto k^{n-3/2}$ instead of $k^{3/2}$.
The power counting on both sides of Eq.~(\ref{iSCBA}) yields a unique solution $n=3$.
Therefore, the Born power law $\Gamma_{\bf k} = \beta k^3$ is not changed
by the self-consistent procedure. The damping coefficient $\beta$
no longer diverges near the decay threshold field because
$\alpha$ drops out from the self-consistent equation for $\Gamma_{\bf k}$.
Instead, $\beta$ exhibits a steplike behavior in $H-H^*$, which is in
accord with the 2D character of the van Hove singularity at the border of
the two-particle continuum.

Another interesting asymptotic regime concerns the close vicinity of
the saturation field with an almost parabolic  magnon dispersion.
As $H\to H_s$, the velocity of the acoustic mode decreases and the
condition of weak nonlinearity applies to a progressively narrower
range of momenta $k \ll 4\sqrt{1-H/H_s}$, reducing  the range of validity
of the asymptotic expression (\ref{Gkasympt}). Outside that ${\bf k}$ domain one can use
the parabolic form for magnon dispersion, $\varepsilon_{\bf k} \approx JSk^2$,
to derive another useful asymptotic expression for the decay rate. In this
regime the decay vertex is a  constant
$V_3^{(1)}({\bf q};{\bf k})  \propto \sqrt{1-H/H_s}$, and the angle
between the emitted and initial magnons can be large, removing restrictions
from the decay phase space. Combining the power-counting arguments
following Eq.~(\ref{naive0}) and performing
analytical integration in Eq.~(\ref{Gamma2_k})  we obtain the
universal momentum-independent expression
\begin{equation}
\Gamma_{\bf k} \approx 16 \left(1-\frac{H}{H_s}\right)  \ ,
\end{equation}
which is valid for $4\sqrt{1-H/H_s}\ll k \ll 1$.

\section{Magnon decay in quantum spin liquids }
\label{sec:spin_liquids}

Magnetic materials with quantum-disordered or spin-liquid-like ground states
have attracted significant attention in the past two decades. The original
idea goes back to \textcite{Anderson73}, who suggested the resonating
valence-bond singlet state as a candidate for a quantum-disordered ground
state of the spin-$\frac{1}{2}$ triangular antiferromagnet. Since then, various types
of spin-liquid ground states have been discussed and classified; see
\textcite{Dagotto96}, \textcite{Balents10}, and \textcite{Lhuillier11}.
On the experimental side, well-known  examples  include  the spin-Peierls compound CuGeO$_3$
\cite{Hase93}, the dimer systems Cs$_3$Cr$_2$Br$_9$ \cite{Leuenberger93}
and TlCuCl$_3$ \cite{Takatsu97}, integer-spin antiferromagnetic chains
\cite{Regnault94}, and many others. A common property of all these materials
is the gap in the excitation spectrum, which separates a spin-singlet ground
state from  the triplet of magnetic excitations with $S=1$. For that reason
they are also referred to as quantum spin-gap magnets. The spin-spin
correlations remain short ranged down to zero temperature with the
correlation length inversely proportional to the gap $\xi \sim 1/\Delta$.
Because of the three fold degeneracy, magnons in quantum spin-gap materials
are often called triplons.

The spin-dimer magnets offer a particularly wide variety of quantum-disordered
states regarding the energy scale of the spin gap and different
effective dimensionalities. They became model systems for studying quantum
phase transitions \cite{Ruegg08,Sachdev08} and the field-induced Bose-Einstein
condensation of magnons \cite{Giamarchi99,Nikuni00}. For an overview of this
subfield of quantum magnetism, see \textcite{Grenier07} and
\textcite{Giamarchi08}. Here we focus on the role of spontaneous
magnon decays in the dynamics of quantum spin-gap magnets. Thereby, we exclude
gapless 1D spin-liquid states with fractionalized spinon excitations as well
as more exotic  forms of spin liquids in higher dimensions that have yet
to find a definitive experimental realization \cite{Balents10}.

The theoretical attention paid to magnon decays is motivated by the inelastic
neutron experiments in two organic quantum spin-gap materials piperazinium
hexachlordicuprate (PHCC) \cite{Stone06} and $\rm (CH_3)_2CHNH_3CuCl_3$
called IPA-CuCl$_3$ \cite{Masuda06,Zheludev07} in which
sharp transformation of triplet excitations upon crossing into the
two-magnon continuum was observed. In PHCC, the magnon lifetime is seen rapidly
decreasing once inside the continuum, whereas in IPA-CuCl$_3$ not even
a trace of the single-particle peak is detected beyond the crossing.
Such an end point was dubbed the termination point of the magnon branch
by analogy with a similar effect in the superfluid $^4$He \cite{StatPhysII}.
The origin of this drastic effect was linked to the nonanalytic behavior of
the magnon self-energy due to two-particle decays by \textcite{Zhitomirsky06}.
The predicted behavior was confirmed by numerical simulations
\cite{Zheng06b,Fischer10} and by an analytical model calculation \cite{Bibikov07}.
In this section we give an overview of theoretical results on two-magnon decays
in quantum-gap antiferromagnets. We discuss a microscopic approach to spin-dimer
magnets based on the bosonic representation of spins in terms of bond operators in
Sec.~\ref{sec:bond_operator}. This is followed by an analysis of the
termination point in the magnon dispersion curve; see Sec.~\ref{sec:qp_breakdown}.
Discussion of the related problem of the decay of longitudinal magnons in weakly ordered
antiferromagnets is included in Sec.~\ref{sec:longitudinal magnons}.

\subsection{Bond-operator formalism}
\label{sec:bond_operator}

\begin{figure}[t]
\centerline{
\includegraphics[width=0.8\columnwidth]{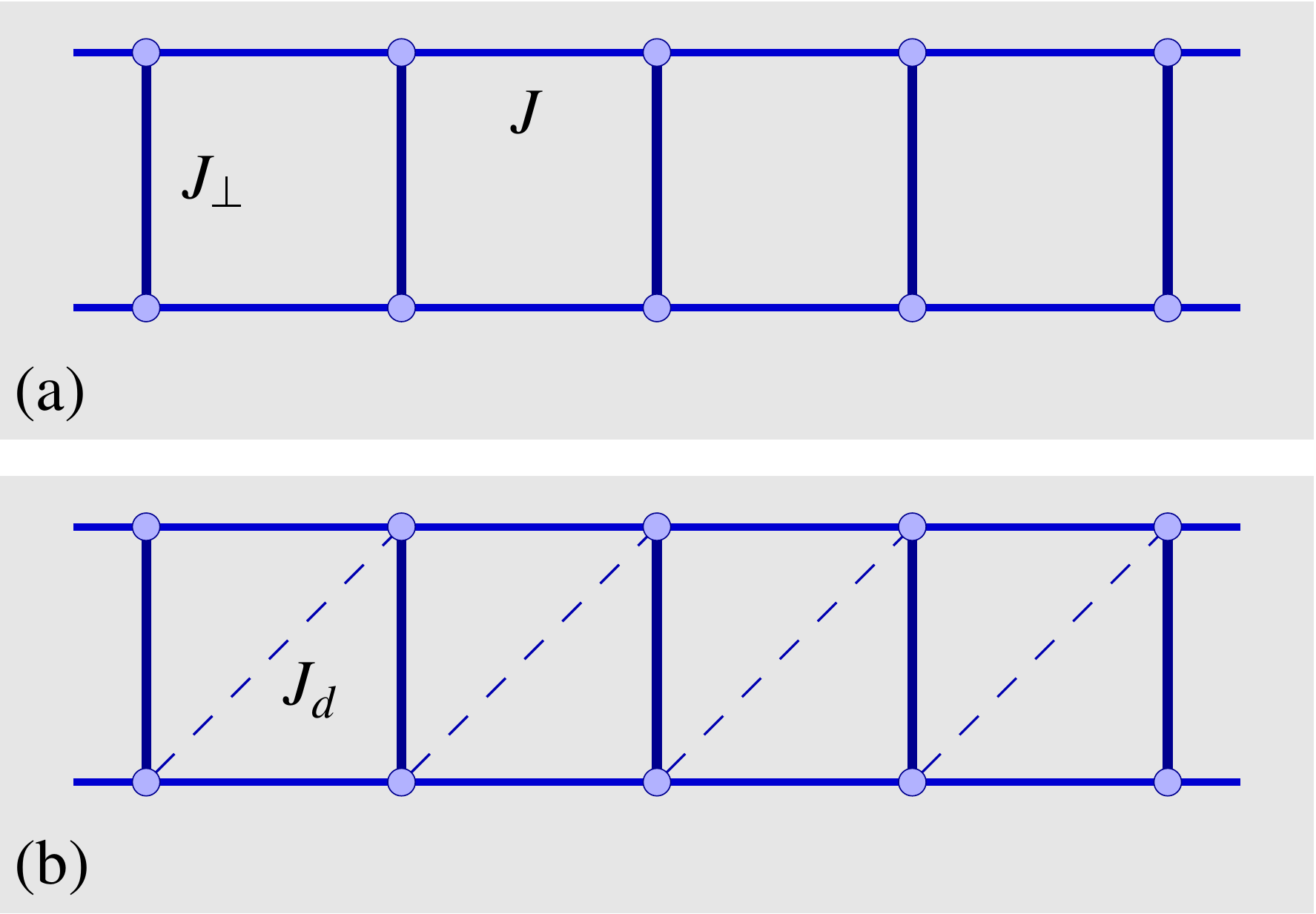}
}
\caption{(color online).\ \
Geometry of exchange bonds for (a) symmetric spin ladder and (b) asymmetric
ladder with extra diagonal coupling.
}
\label{fig:spin_ladder}
\end{figure}

One of the  most studied models of quantum spin-dimer systems is the spin-$\frac{1}{2}$
ladder shown in Fig.~\ref{fig:spin_ladder}(a). Its Hamiltonian is
\begin{equation}
\hat{\cal H} = J_\perp \sum_i {\bf S}_{1i}\cdot{\bf S}_{2i} +
J \sum_{i,n=1,2} {\bf S}_{ni}\cdot{\bf S}_{ni+1}\ ,
\label{Hladder}
\end{equation}
where $J_\perp$ is an antiferromagnetic exchange along rungs and $J$ is a
coupling along the  legs of the ladder. Such a model spin Hamiltonian is realized
in a few magnetic compounds; see, for example, \textcite{Notbohm07} and \textcite{Schmidiger11}.
For $J_\perp \agt J$, the ground state can be approximated by a product of singlet
dimers occupying the rungs of the ladder. In this limit, properties of the spin ladder
can be calculated using the bond-operator representation of spins proposed
independently in several works \cite{Harris73,Chubukov89,Sachdev90}.

Two spins $S=1/2$ on the same rung can form  one singlet $|s\rangle$ and
three triplet states:
\begin{eqnarray}
&& |s\rangle = \frac{1}{\sqrt{2}}(|\!\uparrow\downarrow\rangle - |\!\downarrow\uparrow\rangle)\,,
\quad |t_+\rangle = |\!\uparrow\uparrow\rangle\,,
\nonumber \\
&& |t_0\rangle = \frac{1}{\sqrt{2}}(|\!\uparrow\downarrow\rangle + |\!\downarrow\uparrow\rangle)\,,
\quad |t_-\rangle = |\!\downarrow\downarrow\rangle\,.
\label{st_states}
\end{eqnarray}
In zero field it is more convenient to use the vector basis
$|t_x\rangle = (|t_-\rangle - |t_+\rangle)/\sqrt{2}$,
$|t_y\rangle = i(|t_-\rangle + |t_+\rangle)/\sqrt{2}$, and $|t_z\rangle= |t_0\rangle$;
see \textcite{Sachdev90}. The representation of the original spin operators
in the dimer basis is
\begin{equation}
S_{1,2}^\alpha = \pm \frac{1}{2}\,(s^\dagger t^{_{}}_\alpha+ t_\alpha^\dagger s) -
\frac{i}{2}\, \epsilon^{\alpha\beta\gamma} t_\beta^\dagger t^{_{}}_\gamma \ ,
\label{bond_operator}
\end{equation}
where we introduced singlet $s^\dagger$ and triplet $t_\alpha^\dagger$ operators
that create four physical states out of a fictitious vacuum $|0\rangle$.
Bosonic commutation relations for $s$ and $t_\alpha$ together with
the constraint $s^\dagger s + t_\alpha^\dagger t^{_{}}_\alpha = 1$
ensure the standard SU(2) algebra for spins.

In the bond-operator representation (\ref{bond_operator}), the spin
Hamiltonian (\ref{Hladder}) takes the form
$\hat{\cal H} = \hat{\cal H}_0 + \hat{\cal H}_2 + \hat{\cal H}_4$,
where $\hat{\cal H}_0$ and  $\hat{\cal H}_2$ are quadratic in $t_\alpha$:
\begin{eqnarray}
\hat{\cal H}_0 & = & J_\perp \sum_i \Bigl( -\frac{3}{4}\, s^\dagger_i s^{_{}}_i +
\frac{1}{4}\, t_{\alpha i}^\dagger t^{_{}}_{\alpha i}\Bigr)  \ ,
\\
\hat{\cal H}_2 & = & \frac{J}{2} \sum_i \Bigl(
s_i^\dagger s^{_{}}_{i+1} t_{\alpha i+1 }^\dagger t^{_{}}_{\alpha i} +
s_i^\dagger s_{i+1}^\dagger t^{_{}}_{\alpha i+1 } t^{_{}}_{\alpha i}
+   \textrm{h.c.} \Bigr)  \ ,
\nonumber
\end{eqnarray}
whereas $\hat{\cal H}_4$ includes four-triplon terms. Next one needs to
enforce the local constraint on the total number of singlets and triplets
on each dimer. This can be achieved using various approximate schemes.
One of them consists of condensing singlets  $s_i,s^\dagger_i\to\langle s\rangle$,
and fixing the boson density with the help of the chemical potential \cite{Sachdev90}.
Alternatively, the constraint is resolved in Holstein-Primakoff-like fashion,
 $s = s^\dagger = \sqrt{1- t_{\alpha}^\dagger t^{_{}}_{\alpha}}$, and the square roots
are expanded in powers of the triplon density \cite{Chubukov89}.
Although the final form of the effective Hamiltonian is somewhat different
in the two approaches, the lowest-order magnon-magnon interaction contains only
four-particle terms.

Suppose now that the crystal space group allows additional asymmetric diagonal
coupling $J_d$ between rungs of the ladder [see Fig.\ref{fig:spin_ladder}(b)]:
\begin{equation}
\hat{\cal H}_d = J_d \sum_i {\bf S}_{1i}\cdot{\bf S}_{2i+1} \ .
\label{Haladder}
\end{equation}
The bond-operator representation for $\hat{\cal H}_d$
yields a three-magnon interaction term
\begin{equation}
\hat{\cal H}_3 = \frac{J_d}{2}\langle s\rangle \sum_{k_i}
\epsilon^{\alpha\beta\gamma} \bigl(\sin k_1 \, t_{k_1\alpha}^\dagger t_{k_2\beta}^\dagger
t^{_{}}_{k_3\gamma}\! + \, \textrm{h.c.}\bigr),
\label{H32}
\end{equation}
which has the structure anticipated in Sec.~\ref{sec:int_SL}.

The difference in the magnon interaction between the two models [see
Figs.~\ref{fig:spin_ladder}(a) and (b)] has a symmetry origin.
Three-magnon interactions are prohibited in a symmetric spin ladder
because singlet states are antisymmetric under permutation of two spins
while triplet states have even parity. However, if the reflection symmetry
between the ladder legs is broken, coupling between one- and two-triplet
sectors (\ref{H32}) is also allowed.
Three-magnon interactions exist in many spin-dimer systems
including  the bond-alternating spin chain
and various 2D and 3D arrays  of dimers \cite{Sachdev90,Matsumoto04}.
Cubic terms have also been discussed for the Haldane-chain antiferromagnets
\cite{Kolezhuk06,Zaliznyak01}.
Finally, the geometry of exchange bonds in the spin-dimer magnet IPA-CuCl$_3$,
which will be discussed in the next section, has precisely the structure
shown in  Fig.~\ref{fig:spin_ladder}(b) \cite{Masuda06,Fischer11}.

\subsection{Termination point of the magnon branch}
\label{sec:qp_breakdown}
\begin{figure}[t]
\centerline{
\includegraphics[width=0.95\columnwidth]{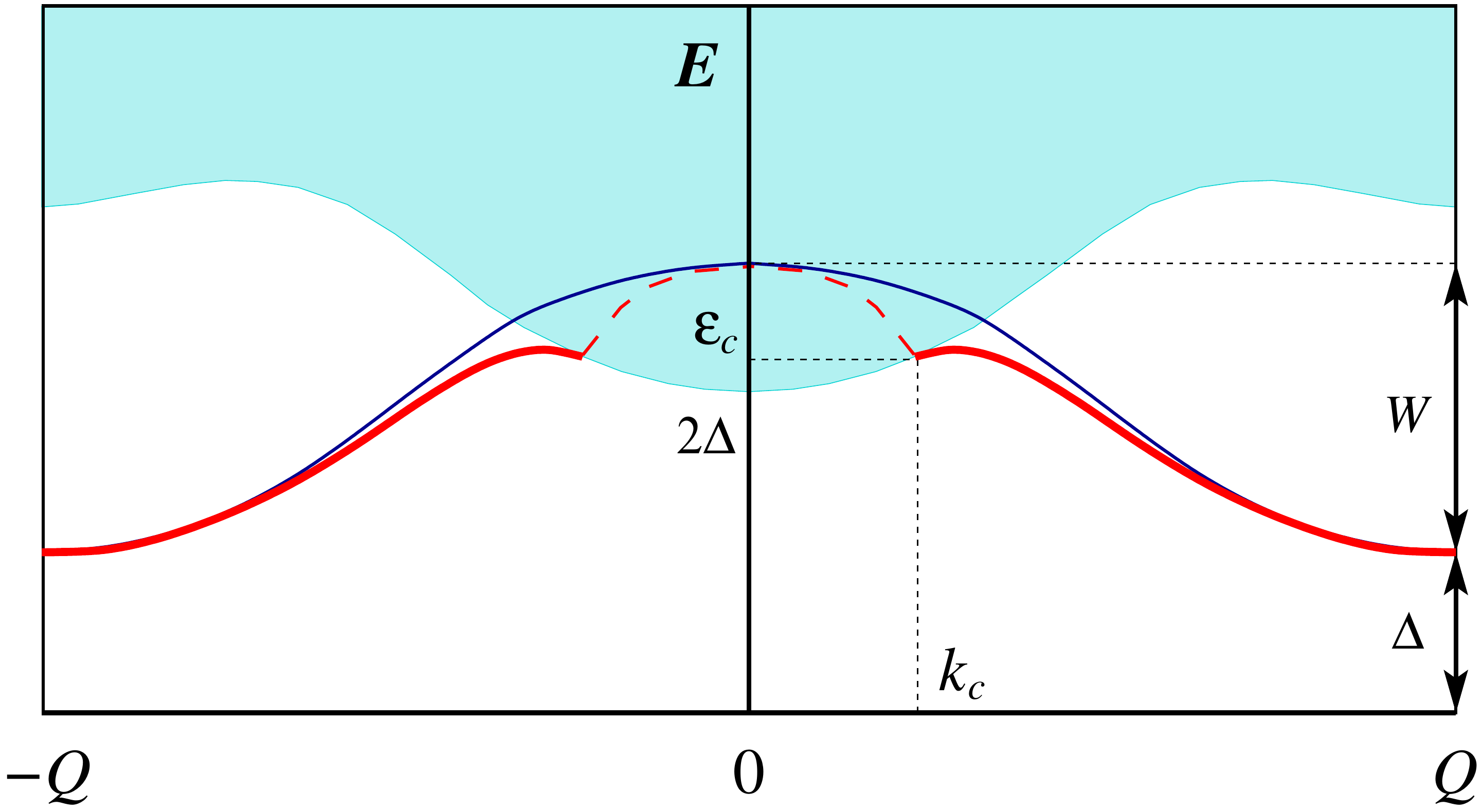}}
\caption{(color online).\ \
Structure of the energy spectrum of a quantum spin-gap magnet.
The thin solid line is the bare triplon energy
$\varepsilon_{\bf k}$, the full solid line is the renormalized dispersion
$\bar\varepsilon_{\bf k}$, and the shaded region is the two-particle continuum.
The decay threshold is marked by $k_c$ and $\varepsilon_c$.
}
\label{fig:crossing}
\end{figure}

We now investigate the effects of three-magnon interactions for the
generic energy spectrum of a quantum spin-gap magnet illustrated in
Fig.~\ref{fig:crossing}. An important feature of the magnon dispersion is
a minimum  at a finite momentum  due to antiferromagnetic  correlations
in the system. For simplicity, we consider only the most common case
of a single minimum at a commensurate wave vector ${\bf Q}$. The overlap
between the single-magnon branch and two-particle continuum occurs if the
magnon bandwidth $W={\cal O}(J)$ exceeds the gap $\Delta={\cal O}(J_\perp)$;
see Fig.~\ref{fig:crossing}. Then, naturally, the decay region is the range
of momenta centered around ${\bf k = 0}$ and magnons that are created in a
decay process are farther down the dispersion curve toward ${\bf Q}$ and $-{\bf Q}$.

Following the kinematic analysis of Sec.~\ref{sec:kinematics_threshold},
in the absence of gapless Goldstone modes, the only two channels that may
define the boundary of the decay region are the decays into magnons with
equal momenta or equal velocities [cases (i) and (iv), respectively].
Considering the typical situation of a modest overlap with the continuum,
 i.e., when the magnon energy $\varepsilon_c \agt 2\Delta$, one can show
that the prevailing channel is the former channel of equal momenta
$\varepsilon_{{\bf k}_c}=2\varepsilon_{({\bf k}_c+{\bf G})/2}$, where
$\bf G$ is a reciprocal lattice vector. The threshold boundary for the
field-induced decays in square-lattice antiferromagnet (see Sec.~\ref{sec:field_decays}),
is of the same type, yet the kinematics is quite different here as the
emitted magnons are outside of the decay region and no cascadelike smearing
of the boundary is possible. Thus, we encounter a kinematic situation distinct
from the cases treated in Secs.~\ref{sec:field_decays} and \ref{sec:tafm},
which requires detailed consideration of the spectrum near the decay boundary.
As shown next, the presence of nonanalytic self-energy leads to dramatic
qualitative changes of the magnon spectrum near the threshold.

According to the preceding discussion, the three-magnon interaction has the form
\begin{equation}
\hat{V}_3 = \frac{1}{2}\! \sum_{{\bf k},{\bf q}} V_3({\bf q},{\bf k-q};{\bf k})
\epsilon^{\alpha\beta\gamma}\bigl(t^\dagger_{{\bf q}\alpha}
t^\dagger_{{\bf k-q}\beta}t^{_{}}_{{\bf k}\gamma }\! + \textrm{h.c.}\bigr).
\label{decayV}
\end{equation}
Although due to conservation of the total spin the magnon with  polarization
$\gamma$ can decay only into magnons with different polarizations,
as enforced by the antisymmetric tensor $\epsilon^{\alpha\beta\gamma}$, this
has no implications for the decays as long as the three branches remain degenerate.
More important is the antisymmetric structure of
the magnon vertex under the permutation of polarizations
$\alpha$ and $\beta$ and, consequently, under the interchange of momenta of
the decay products $V_3({\bf q},{\bf q}';{\bf k})=-V_3({\bf q}',{\bf q};{\bf k})$.
Since at the decay boundary the emitted magnons have equal momenta,
the three-magnon  vertex (\ref{decayV}) vanishes at the threshold $V_3({\bf q}^*,{\bf q}^*; {\bf k}_c)=0$.
Then the main difference from the analysis of the decay singularities in
Sec.~\ref{sec:kinematics_singular} is the effective increase of dimensionality due to further
expansion of the decay vertex in $\tilde{\bf q}= {\bf q}-{\bf q}^*$ needed in the vicinity of the boundary:
\begin{equation}
\Sigma({\bf k},\omega)\simeq - g_3^2 \int_0^\Lambda \frac{\tilde{q}^{D+1}\,\textrm{d}\tilde{q}}
{A\tilde{q}^2 + {\bf v}_2\!\cdot\!\Delta{\bf k}-\Delta\omega -i0} \ ,
\label{SelfE_singular}
\end{equation}
where $D$ is the dimension, $\Delta\omega = \omega - \varepsilon_c$,
$\Delta{\bf k} = {\bf k} - {\bf k}_c$,  ${\bf v}_2$ is the velocity of
emitted magnons, $A={\cal O}(J)$ is a constant, and $g_3={\cal O}(J_d)$ is an effective coupling constant.
Because of the vanishing vertex, the self-energy (\ref{SelfE_singular})
is not singular at the decay boundary  $\Delta\omega,|\Delta{\bf k}| \rightarrow 0$ even in $D=1$,
but it is still  nonanalytic.
The higher-order corrections can be shown to leave unchanged the functional form of the nonanalyticity
obtained in the one-loop approximation (\ref{SelfE_singular}).
Next we absorb the regular part of $\Sigma({\bf k},\omega)$
into the magnon energy $\varepsilon_{\bf k}$, while the renormalized spectrum
will be  obtained from the Dyson equation with the nonanalytic part  $\tilde\Sigma({\bf k},\omega)$.
We focus on the 1D case, mentioning a few results for the higher-dimensional systems  in
the end.

We begin with the region  $k>k_c$ where spontaneous magnon decays are forbidden;
see Fig.~\ref{fig:crossing}.
In this case,    integration in  Eq.~(\ref{SelfE_singular}) yields the following nonanalytic self-energy:
\begin{equation}
\tilde{\Sigma}(k,\omega) =
\lambda \sqrt{v_2\Delta k -\Delta\omega} \ , \qquad
\Delta\omega<v_2\Delta k
\label{SigmaR1d}
\end{equation}
with $\lambda = \pi g_3^2/2A^{3/2}={\cal O}(J_d^2/J^{3/2})$.  Close to the decay boundary
the nonanalytic $\tilde{\Sigma}(k,\omega)$
dominates the regular part of $G^{-1}(k,\omega)\sim\Delta k,\Delta\omega$ and completely
determines the behavior of the quasiparticle pole. Solving the Dyson equation with Eq.~(\ref{SigmaR1d})
yields the renormalized spectrum sketched in Fig.~\ref{fig:crossing}
\cite{Zhitomirsky06}.
Far away from the decay boundary
the renormalized magnon energy $\bar{\varepsilon}_k$ follows
 bare dispersion $\varepsilon_k$ with the negative velocity
$v_1=d\varepsilon_k/dk<0$. On the other hand,
for $k\rightarrow k_c$  the renormalized slope must change its sign to
$d\bar{\varepsilon}_k/dk = v_2>0$ as the minimum of the two-particle
continuum at $k_c$ is formed by magnons  from the positive slope of $\varepsilon_k$.
This change of the magnon velocity takes place in the
crossover region $\Delta k\sim\lambda^2/(v_2-v_1)={\cal O}(J_d^2/J^2)$
and the overall shape of $\varepsilon_k$ resembles the avoided crossing
between the magnon branch $\bar{\varepsilon}_k$  and  the continuum boundary.
However, due to the nonanalytic nature of the self-energy
the behavior of the spectrum is markedly different from the avoided crossing:
 the quasiparticle weight $Z_k$ changes drastically in the
crossover regime and vanishes at the intersection of the magnon branch with the two-magnon continuum.
Such a suppression of the quasiparticle peak
is a hallmark of the termination point in the energy  spectrum,
very similar to that in $^4$He \cite{Pitaevskii59}.
In a different context, similar distortion of the excitation curve $\varepsilon_k$ in the vicinity of the
continuum boundary was also observed by \textcite{Coldea10}.

For $k<k_c$, inside the continuum where spontaneous decays are
allowed,  the nonanalytic self-energy (\ref{SelfE_singular}) is now purely
imaginary:
\begin{equation}
\tilde{\Sigma}(k,\omega) = -i
\lambda \sqrt{|v_2\Delta k -\Delta\omega|}\,, \quad
\Delta\omega>v_2\Delta k\,.
\label{SigmaI1d}
\end{equation}
Within the on-shell approximation, $\omega=\varepsilon_k$, this leads to the following
 magnon decay rate:
\begin{equation}
\Gamma_k \simeq \lambda \sqrt{(v_2-v_1)|k-k_c|} \ .
\label{lifetime1}
\end{equation}
The same result was  obtained by \textcite{Kolezhuk06}.
A more rigorous off-shell solution of the Dyson equation in the complex plane
changes the perturbative result of Eq.~(\ref{lifetime1}) in the $\Delta k$ vicinity  [$\sim{\cal O}(J_d^2/J^2)$]
of the decay boundary to $\Gamma_k \propto |k-k_c|^{2/3}$. However, more importantly,
the quasiparticle weight remains suppressed in this region, in the same way as on the other side of the
termination point. The quasiparticle peak
may reappear farther away from the decay boundary depending on the strength of the three-magnon interaction.

We also note  the common misconception regarding the absence
of a pole beyond the termination point inside the two-particle continuum \cite{Stone06}.
According to our analysis, such a pole is always present on either side of the termination
point, but its quasiparticle residue is vanishingly small in the vicinity of that point.

The effect of spontaneous decays in the energy spectra of gapped spin-liquids
was studied numerically for the bond-alternating spin-$\frac{1}{2}$ chain by \textcite{Zheng06b}
and for the asymmetric ladder by \textcite{Fischer10}.
For both models, the single-particle branch merges with the lower edge of the continuum,
gradually losing its spectral weight. \textcite{Fischer10} also found some indications
of a reemerging quasiparticle peak inside the continuum for small values of
the three-magnon coupling $J_d$.

\begin{figure}[t]
\centerline{
\includegraphics[width=0.7\columnwidth]{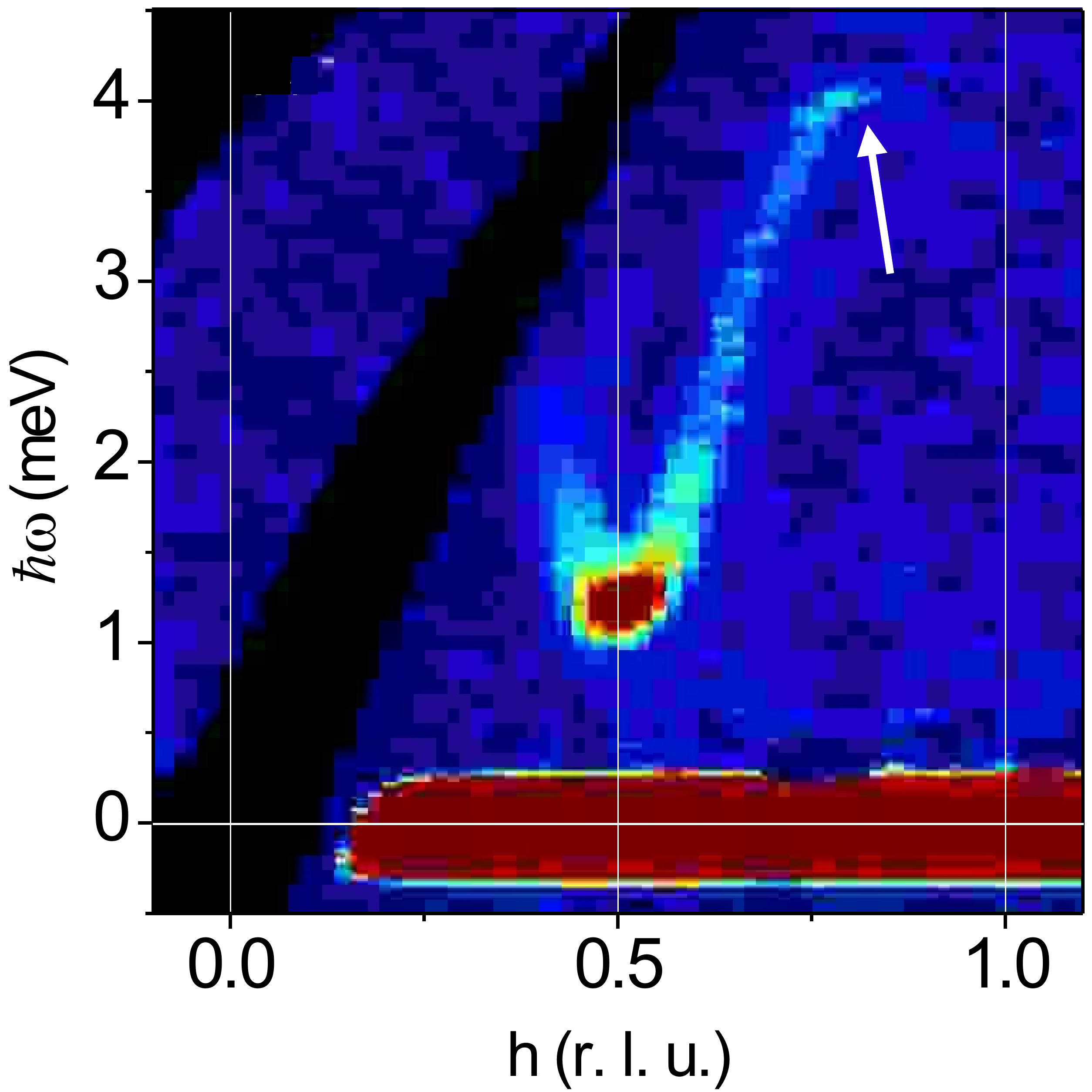}}
\caption{(color online).
Time-of-flight inelastic neutron-scattering data measured
in IPA-CuCl$_3$ at $T=1.5$~K \cite{Masuda06}. The arrow indicates
the termination point of the magnon branch. From Andrey Zheludev.
}
\label{fig:ipa}
\end{figure}

A remarkable illustration of the theory
is provided by the energy spectra measured in the spin-ladder
compound IPA-CuCl$_3$  \cite{Masuda06}. This
material consists of 2D arrays of weakly coupled
asymmetric $S=1/2$ Heisenberg ladders [see Fig. \ref{fig:spin_ladder}(b)],
with the coupling constants
$J_\perp = 2.9$~meV, $J = 1.2$~meV,  $J_d = -2.3$~meV, and
a small interladder exchange $J_{2D}\approx -0.3$~meV
\cite{Fischer11}. The resolution-limited peaks corresponding to the
magnon band with a minimum at $Q=0.5$ (units of $2\pi/a$) and the
energy gap $\Delta=1.2$~meV are observed in IPA-CuCl$_3$; see Fig.~\ref{fig:ipa}.
However, the magnon branch abruptly  disappears at $k_c\approx 0.8$.
Using their data for $\varepsilon_{\bf k}$,
\textcite{Masuda06} determined that $k_c$ indeed corresponds
to the intersection point of $\varepsilon_{\bf k}$ with the continuum.
Thus, the single-magnon branch in IPA-CuCl$_3$  terminates
at the boundary of the two-magnon continuum, as in
the superfluid $^4$He, where the quasiparticle branch terminates
at the threshold of two-roton decays \cite{Glyde98,Pitaevskii59}.

A gradual broadening of the quasiparticle peak by two-magnon decays without
its complete disappearance was observed in the 2D dimer organic material
PHCC \cite{Stone06} and in the quasi-1D antiferromagnet $\rm CsNiCl_3$
\cite{Zaliznyak01}. The difference from IPA-CuCl$_3$ discussed above is mainly
due to the higher dimensionality of these two materials. In the case of PHCC,
the experimental data are reasonably well described by Fermi's golden-rule
formula
\begin{equation}
\Gamma_{\bf k}^{(D)} \simeq \lambda |{\bf k}-{\bf k}_c|^{D/2} \ ,
\label{lifetimeD}
\end{equation}
with $D=2$ \cite{Kolezhuk06}. Beyond the on-shell approximation, the triplon
self-energy (\ref{SelfE_singular}) in $D>1$ gives much weaker nonanalyticity.
For example, in 2D,
$\tilde{\Sigma}({\bf k},\omega) \sim \Delta\omega\ln(\Lambda/\Delta\omega)$,
and the suppression of  the quasiparticle peak in PHCC should occur only in
an exponentially small region near the crossing point, justifying the validity
of the perturbative result (\ref{lifetimeD}).

Thus far, we discussed spontaneous magnon decays in quantum spin-gap systems
under the ``standard'' condition of  the cubic
vertex vanishing at the decay threshold boundary.
The decay-related singularities may be substantially enhanced, however,
if the vertex is finite. This takes place, for example, when the threshold
channel  corresponds to  decays into magnons
with different momenta. Another mechanism is spin anisotropy, which
violates spin conservation and removes the odd parity of the cubic vertex (\ref{decayV}) with respect
to spin indices and momenta, making the decay vertex finite at the continuum boundary and also
lifting degeneracy of the triplon bands. This mechanism might be relevant to
the spin-1 alternating chain antiferromagnet Ni(C$_9$H$_{24}$N$_4$)(NO$_2$)ClO$_4$ (called NTENP), whose
unusual spin dynamics in external magnetic field may signify
spontaneous two-magnon decays \cite{Hagiwara05,Regnault06}.
Finite magnetic fields also produce intriguing changes in the dynamical
properties of IPA-CuCl$_3$ \cite{Garlea07,Zheludev07}.
Theoretical explanation of these experimental findings remains an open problem.

Finally we note that the occurrence of the terminationlike behavior in the spectra
of quantum spin-gap magnets  is by no means unique. It can also take place in
a variety of spin-anisotropic antiferromagnets \cite{Chernyshev09a}, in which the
anisotropy gap stabilizes the decay products and makes
the corresponding threshold singularities essential.

\subsection{Decay of longitudinal magnons}
\label{sec:longitudinal magnons}

Our final example of spontaneous magnon decays concerns weakly-ordered
collinear antiferromagnets, which occupy an intermediate place between
quantum-disordered and well-ordered semiclassical magnets.
Apart from the conventional spin waves, excitation spectrum of the weakly-ordered
spin system may display a longitudinal mode that  corresponds to fluctuations
of the magnitude of the order parameter \cite{Affleck89}. The presence of this
mode can be easily understood by considering a transition  from the spin-liquid phase
into an ordered state under an external pressure $P$ \cite{Matsumoto04}.
In the quantum-disordered state, up to the critical pressure $P_c$
where the gap closes, three  magnon modes are degenerate. Above the quantum transition,
$P>P_c$, two of these modes transform into the Goldstone modes of the ordered magnetic
structure, while the remaining third excitation acquires a finite gap.
At this point one may ask how the above scenario could at all agree with the
conventional spin-wave picture, which accounts only for the two
magnon modes. The apparent contradiction is resolved if we notice that
the longitudinal mode is always unstable with respect to decay into a pair
of transverse spin waves and thus has an intrinsic linewidth \cite{Affleck92}.
Farther inside the parameter domain of the ordered phase, the longitudinal excitation
transforms into two-magnon resonance and eventually disappears.

The pressure-induced quantum critical point is realized, for instance,
in the spin-dimer antiferromagnet $\rm TlCuCl_3$ \cite{Oosawa03}.
The low critical pressure $P_c\simeq 1$~kbar  enables the use of inelastic neutron measurements
to explore  the phases on both sides of the critical
point. The existence of the gapped longitudinal mode in the antiferromagnetic phase
of $\rm TlCuCl_3$ was demonstrated by
\textcite{Ruegg08}, who concluded that such a mode is ``critically well defined''
because of its large broadening.
Recently, the lifetime of
longitudinal excitations in $\rm TlCuCl_3$  was calculated by
\textcite{Kulik11}. In the following, we describe basic
aspects of spontaneous magnon decays in weakly ordered collinear
antiferromagnets using the example of $\rm TlCuCl_3$.

The harmonic theory of magnetic excitations in both disordered and ordered
phases of $\rm TlCuCl_3$ was developed by \textcite{Matsumoto04} who used
the bond-operator formalism. At the critical pressure
$P=P_c$ the gap of the triplon branch closes at the commensurate ordering wave vector
$\bf Q$. No spontaneous decays of the low-energy magnons exist
up to this point. At $P>P_c$  one member of the degenerate triplet condenses,
$t_{iz}\to \langle t_z\rangle e^{i{\bf Q}\cdot {\bf r}_i}$, and the collinear
magnetic structure develops along the corresponding direction.
The degeneracy of the triplon bands  is removed in the ordered
phase. Transverse excitations $t_{x,y}$ stay degenerate with each other and remain gapless
with the energy $\varepsilon^\perp_{\bf Q+k} \approx ck$,
whereas the fluctuating part of the condensed magnon $\tilde{t}_z=t_z-\langle t_z\rangle$
represents the longitudinal mode with the energy $\varepsilon^z_{\bf k}$ and the gap
$\Delta_L = \varepsilon^z_{\bf Q}\propto \sqrt{P-P_c}$.

Although  the cubic vertex (\ref{decayV}) may be absent in the disordered state
at $P<P_c$, the triplet condensation in the ordered phase always produces a
three-particle interaction involving
longitudinal fluctuations; see Eq.~(\ref{condensate}):
$t^\dagger_{\alpha{\bf k}_4}t^\dagger_{\alpha{\bf k}_3}t^{_{}}_{\beta{\bf k}_2}t^{_{}}_{\beta{\bf k}_1}
\to \langle t_z\rangle \tilde{t}^\dagger_{z{\bf k}_3}t^{_{}}_{\beta{\bf k}_2}t^{_{}}_{\beta{\bf k}_1}$.
Thus, the symmetry ensures that the longitudinal magnon can decay  into a pair of transverse
magnons  $\beta=x,y$.  Both energy and momentum conservation are also trivially satisfied
for the decay of the gapped mode in the vicinity of the ordering wave vector $\bf Q$.
The asymptotic form of the decay vertex  is $V_3({\bf q},{\bf q}';{\bf k}) \simeq \langle t_z\rangle/
(\varepsilon^z_{\bf k}\varepsilon^\perp_{\bf q} \varepsilon^\perp_{\bf q'})^{1/2}$ \cite{Kulik11}.
In the bond-operator formalism, the  anomalous denominator
 in $V_3$ with $\varepsilon^\perp_{\bf Q+q}\approx cq$ appears from  the coefficients
of the Bogolyubov transformation. The decay rate of the longitudinal
magnon  at $\bf k=Q$ calculated from  Eq.~(\ref{Gamma2_k}) is
\begin{equation}
\Gamma_{\bf Q}  \simeq \frac{\langle t_z\rangle^2}{\Delta_L} = \gamma \Delta_L\ ,
\label{GammaL}
\end{equation}
where the final relation follows from the mean-field result for the expectation value of the condensate
fraction $\langle t_z\rangle \propto \sqrt{P-P_c}$.
The theoretical estimate of the damping coefficient $\gamma=0.2$ is not far from the experimental value
$\gamma\approx 0.15$ and can be further improved upon  by taking into account spin anisotropy in
$\rm TlCuCl_3$ \cite{Kulik11}. Thus, the decay rate of the longitudinal mode is only  parametrically
small compared to its energy and whether it remains well-defined  depends on
microscopic parameters of the system.

\section{Summary and outlook}
\label{sec:summary}

We presented an overview of the phenomenon of spontaneous magnon decays
in quantum antiferromagnets and discussed symmetry and kinematic conditions
that are necessary for it. While no exhaustive list of magnets that favor
quasiparticle decays can be given, the offered examples outline the principal
classes of spin systems, in which the effect may occur. Generally, the noncollinear
spin structures in the systems where the symmetry is broken spontaneously from
the high-symmetry spin Hamiltonian belong to one of such classes. The gapped
quantum spin liquids is yet another class and a proximity to a quantum critical
point can be another common situation in which magnetic excitations may be
intrinsically damped.

The hallmarks of the decay-induced anomalies in magnon spectra include substantial
broadening of magnon peaks in a large part of the Brillouin zone, termination points, and
non-Lorentzian features in the structure factor,  as well as strong deviations of the
spectrum from the harmonic theory, especially for $S=1/2$ systems. With improvement
in the resolution of the neutron-scattering experiments expected in the near future,
these effects should become readily observable. Besides their effect on the dynamical
properties, magnon decays may also play a  role in the critical behavior in the vicinity
of quantum critical points \cite{Fritz11}.

We also highlighted the role of singularities, which are generally concomitant with
decays and are related to the crossing of the single-particle branch with the surface of
van Hove singularities in the two-particle continuum. Regularization of such singularities
represents a theoretical challenge that is largely unfamiliar in the traditional spin-wave
theory and requires the use of self-consistent schemes exemplified in this work. While
most of the systems considered here are 1D or 2D, for which the effects of spontaneous
magnon decays are amplified, the strong effects due to spontaneous decays are not restricted
to   low-dimensional systems, as shown in Sec.~\ref{sec:field_decays}.

The direct experimental evidence of magnon decays in low-dimensional spin-dimer systems
discussed in Sec.~\ref{sec:spin_liquids} is unequivocal and progress is being made
toward similar observations in the noncollinear and field-induced settings. For the
latter case,  several materials with sufficiently small interactions that allow magnetic
fields to fit into the window accessible by neutron-scattering experiments have
recently been synthesized and the first indication of  field-induced magnon decays has been
found in the spin-$\frac{5}{2}$ square-lattice antiferromagnet $\rm Ba_2MnGe_2O_7$ \cite{Masuda10}.
As noted in Sec.~\ref{sec:field_decays}, the BEC magnets are also very promising
for the observation of the field-induced decays in the vicinity of their lower quantum-critical
point, which is readily accessible in a number of them. The experimental search for spontaneous
magnon decays in noncollinear magnets is also ongoing, with several $S=1/2$ triangular-lattice
antiferromagnets, such as $\rm Cs_2CuBr_4$ \cite{Tsujii07} and $\rm Ba_3CoSb_2O_9$ \cite{Shirata12},
and also the large-$S$ weakly anisotropic noncollinear antiferromagnets \cite{Ishii11,Toth11}
being most promising.

Besides improvements in conventional neutron scattering, there has been
a recent dramatic improvement in neutron-spin-echo  spectroscopy technique;
see \textcite{Neutron_Spin_Echo}. The energy resolution of this method can reach the $\mu$eV
range compared to the meV resolution for triple-axis spectroscopy. Successful
measurements of magnon lifetimes by the neutron-spin-echo technique  were reported
by \textcite{Bayrakci06}, \textcite{Nafradi11}, and \textcite{Chernyshev12a} and more
of such studies are forthcoming. This development may  offer an opportunity to observe
a variety of new physical effects including spontaneous magnon decays.

Finite magnon lifetimes can also be detected by experimental probes other than
the neutron scattering. One such example is the decay-related smearing observed
by electron spin resonance in the bond-alternating chain antiferromagnet NTENP
\cite{Glazkov10}. Another promising experimental technique is resonant inelastic
x-ray scattering \cite{Ament11}, which is able to detect excitation broadening in
large-exchange spin systems. The magnetic component of thermal conductivity should
also be strongly influenced by spontaneous magnon decays \cite{Kohama2011}.

\begin{acknowledgments}
We thank our co-workers, W. Fuhrman, M. Mourigal, and V. Stephanovich,
for their contributions to this work and O. Sylju\aa sen and A. Zheludev for providing
figures for this article. We are grateful to I. Affleck, C. Batista, W. Brenig, A. Chubukov,
R. Coldea, F. Essler, N. Hasselmann, G. Jackeli, E. Kats, P. Kopietz, B. Lake, A. L\"auchli,
A. Muratov, B. Normand, N. Perkins, L.-P. Regnault, H. R\o{}nnow, Ch.\ R\"uegg, O. Starykh,
O. Sushkov, I. Zaliznyak, and A. Zheludev for insightful discussions. This work was supported
by the DOE under Grant No.~DE-FG02-04ER46174 (A. L. C.).
\end{acknowledgments}

\bibliographystyle{apsrmp}

\end{document}